\documentclass[11pt,letterpaper]{article}
\pdfoutput=1
\usepackage{jheppub}
\usepackage{color}
\usepackage{graphicx}

\usepackage{verbatim}
\usepackage{amsmath}
\usepackage{amssymb}
\usepackage{subfig}
\usepackage{url}
\usepackage{bbold}
\usepackage{slashed}

\usepackage{multirow}
\usepackage{threeparttable}
\usepackage{paralist}

\DeclareRobustCommand{\Sec}[1]{Sec.~\ref{#1}}
\DeclareRobustCommand{\Secs}[2]{Secs.~\ref{#1} and \ref{#2}}

\DeclareRobustCommand{\Fig}[1]{Fig.~\ref{#1}}

\DeclareRobustCommand{\Eq}[1]{Eq.~(\ref{#1})}

\DeclareRobustCommand{\EqsOr}[2]{Eqs.~(\ref{#1}) or (\ref{#2})}

\newcommand{\para}{\paragraph{}}
\newcommand{\be}{\begin{equation}}
\newcommand{\ee}{\end{equation}}

\newcommand{\ev}{{\text{event}}}

\newcommand{\gev}{\mathrm{GeV}}
\newcommand{\tev}{\mathrm{TeV}}
\newcommand{\MET} {E_T^{{\text{miss}}}}
\newcommand{\METx}{E_x^{\text{miss}}}

\newcommand{\pt}{$p_T$~}

\newcommand{\meanf}{\bar{\alpha}^{F}_\text{PU}}

\newcommand{\rmsf}{\sigma^{F}_\text{PU}}

\begin{document}
\title{Pileup Per Particle Identification}

\author[a]{Daniele Bertolini,}
\author[b]{Philip Harris,}
\author[c]{Matthew Low,}
\author[d]{and Nhan Tran}

\affiliation[a]{Center for Theoretical Physics, Massachusetts Institute of Technology, Cambridge, MA 02139, USA}
\affiliation[b]{CERN, European Organization for Nuclear Research, Geneva, Switzerland}
\affiliation[c]{Enrico Fermi Institute and Kavli Institute for Cosmological Physics, University of Chicago, Chicago, IL 60637, USA}
\affiliation[d]{Fermi National Accelerator Laboratory (FNAL), Batavia, IL 60510, USA}

\emailAdd{danbert@mit.edu}
\emailAdd{philip.coleman.harris@cern.ch}
\emailAdd{mattlow@uchicago.edu}
\emailAdd{ntran@fnal.gov}

\abstract{
We propose a new method for pileup mitigation by implementing ``pileup per particle identification" ({\tt PUPPI}).  For each particle we first define a local shape $\alpha$ which probes the collinear versus soft diffuse structure in the neighborhood of the particle.  The former is indicative of particles originating from the hard scatter and the latter of particles originating from pileup interactions.  The distribution of $\alpha$ for charged pileup, assumed as a proxy for all pileup, is used on an event-by-event basis to calculate a weight for each particle. The weights describe the degree to which particles are pileup-like and are used to rescale their four-momenta, superseding the need for jet-based corrections.  Furthermore, the algorithm flexibly allows combination with other, possibly experimental, probabilistic information associated with particles such as vertexing and timing performance.  We demonstrate the algorithm improves over existing methods by looking at jet $p_T$ and jet mass. We also find an improvement on non-jet quantities like missing transverse energy.
}

\preprint{
\begin{flushright}
  EFI {14-18}  \\ FERMILAB-PUB-14-238-PPD \\ MIT-CTP {4558}
\end{flushright}
}
\maketitle

\section{Introduction}
\label{sec:introduction}

Pileup, {\it i.e.} overlapping secondary proton-proton collisions on top of the primary interaction, will be a major challenge for the high luminosity LHC runs.  Several methods for dealing with pileup are being successfully applied by ATLAS~\cite{Aad:2013gja,ATLAS-CONF-2013-083,ATLAS-CONF-2013-085,CMS:2013wea,ATLAS-CONF-2014-018,ATLAS-CONF-2014-019} and CMS~\cite{Chatrchyan:2011ds,Chatrchyan:2013vbb,CMS-PAS-JME-12-002} on present data.  Current methods, however, will be strained both in upcoming LHC runs with expected pileup levels of $n_{\text{PU}}=140$ or more, and at possible future hadron colliders.  Recently, newer ideas for pileup mitigation have been proposed.  A brief summary of the state-of-the-art is given below:

\begin{itemize}
  \item {\it Four-vector area subtraction}~\cite{Cacciari:2007fd,Cacciari:2008gn}: corrects the four-vector of a jet based on the characteristic pileup density of an event and on the jet area.  It has been applied by ATLAS and CMS, but requires additional experimental tuning on top of area $\times$ pileup density subtraction~\cite{Chatrchyan:2011ds,ATLAS-CONF-2013-083}.

  \item {\it Shape subtraction}~\cite{Soyez:2012hv}: a generalization of area subtraction from the jet four-vector to jet shapes, {\it e.g.} jet mass.  Each shape is separately corrected using the same pileup density measure as area subtraction and using the susceptibility of a given shape to soft uniform contamination.

  \item {\it Grooming ({\it e.g.} filtering~{\normalfont\cite{Butterworth:2008iy}}, pruning~{\normalfont\cite{Ellis:2009me,Ellis:2009su}}, trimming~{\normalfont\cite{Krohn:2009th}}, soft drop~{\normalfont\cite{Larkoski:2014wba}})}: systematically discards a subset of a jet's constituents that are believed to obscure the signal process.  Grooming can be used with subtraction. 

  \item {\it Pileup jet identification}~\cite{ATLAS-CONF-2013-083,CMS:2013wea}: removes entire jets that are identified as being composed primarily of pileup using both charged particle information and jet shapes.

  \item {\it Topological clustering}~\cite{Aad:2011he,Lampl:2008zz,Barillari:2009zza}: calorimeter cell signals are required to be several standard deviations above the typical noise level in the cells.  These high significance cells are used as seeds to form local cell clusters used as inputs to jet algorithms.

  \item {\it Charged hadron subtraction (CHS)}~\cite{CMS:2009nxa}: removes charged particles from pileup based on the vertex from which particles originate.  Four-vector area subtraction is then applied using the remaining particles.
  
  \item {\it Cleansing\footnote{Depending on the precise definition of grooming, this may also be considered a groomer.}}~\cite{Krohn:2013lba}: uses vertex information to remove charged pileup and rescales neutral pileup based on charged pileup composition in a subjet.

  \item {\it Constituent subtraction}~\cite{CMS-DP-2013-018,Berta:2014eza}: extends four-vector area subtraction and shape subtraction to particle level by representing the event density $\rho$ with an area assigned to each particle, correcting the particles's four-vector.
    
\end{itemize}

The methods listed above progressively move from a global approach towards a more local one.  We note that, broadly speaking, the methods utilize three basic pieces of information to identify pileup: the event-wide pileup density, vertex information from charged tracks, and the local distribution of pileup with respect to particles from the leading vertex.  As each technique has advantages and disadvantages it is unlikely that a single method alone will optimally remove pileup.  It is therefore crucial to have a flexible framework to integrate the various pieces of information.  We propose an algorithm to combine this information.  This method uses both global information from the event, as well as local information to identify pileup at the particle level.  As a shorthand, we refer to the method as {\tt PUPPI} (PileUp Per Particle Identification).  The algorithm is intended to remove pileup, rather than just correct jet quantities, to ultimately produce a consistent event interpretation.

It has been shown~\cite{Krohn:2013lba} that individually rescaling the four-momenta of particles in a jet ({\it i.e.} the jet's constituents) is useful not only for correcting kinematic variables, but also for correcting jet shapes in an observable-independent way.  Following the ``jets without jets'' paradigm \cite{Bertolini:2013iqa}, we propose a local approach, in which no clustering is performed and a weight is assigned to each individual particle.  We then choose to use the weight to rescale the particle four-momentum. Ideally, particles coming from pileup would get a weight of zero and particles coming from the hard scatter would get a weight of one. This leads to a pileup-corrected event, where one can then proceed with jet finding without the need for further pileup correction.  In fact, given the pileup-corrected event, event shapes can be measured with a reduced sensitivity to pileup. We show results for jet $p_T$, jet mass, and missing transverse energy and demonstrate that our algorithm improves over existing methods. We find the improvements on missing  transverse energy reconstruction particularly relevant, as disentangling pileup contamination from missing energy is typically harder than for jet-based observables. 

We anticipate that the {\tt PUPPI} algorithm could potentially improve pileup mitigation for other jet and event shapes, as well as the identification of isolated photons and leptons. More generally, such a {\it per particle} approach may contribute valuable input into the design of future detectors by highlighting the complementary information measured by several detector subsystems.

The rest of the paper is organized as follows: in \Sec{sec:algorithm} we describe the algorithm, in \Sec{sec:setup} we describe the setup we used for generating Monte Carlo data, and in \Sec{sec:results} we present our results. Finally, we conclude and discuss future work in \Sec{sec:summary}.

\section{The Algorithm}
\label{sec:algorithm}

Before discussing details, we describe qualitatively how the algorithm works.  First we select a shape $\alpha$, which attempts to locally distinguish parton shower-like radiation from pileup-like radiation, then compute it for each particle in an event.  A basic handle to distinguish pileup and leading vertex particles is given by the $p_T$ spectrum, with the pileup spectrum falling much faster.  While we make use of this feature in the algorithm, the shape $\alpha$ itself attempts to exploit additional and complementary information with respect to the $p_T$ of a single particle, as discussed in \Secs{sec:LocalShape}{sec:metric}.  Where tracking is available, we know the answer to whether charged particles are from the leading vertex (LV) or from a pileup vertex (PU).  We can use the median and the RMS of the $\alpha$ values for charged pileup as an event-level characterization of the pileup distribution.

Next we assign a weight to each particle by comparing its $\alpha$ value to the median of the charged pileup distribution.  The weight takes values between zero and one and indicates how much a particle is allowed to contribute to an event.  Ideally, particles from the hard scatter would get a weight of one and pileup particles would get a weight of zero.  Almost all pileup particles have $\alpha$ values within a few standard deviations of the median and are assigned small weights.  On the other hand, $\alpha$ values that deviate far from the charged pileup are very uncharacteristic of pileup, and these particles are assigned large weights.  As discussed in \Sec{sec:addninfo}, our method for computing weights allows for experimental information, such as vertexing and timing performance, to be smoothly incorporated.

Finally, we choose to apply the weights to rescale the particle's four-momentum.
Particles with a very small weight or with a very small rescaled $p_T$ are discarded.  The final set of pileup-corrected particles can now be used as input to a jet algorithm or directly in the calculation of missing energy.

The rest of this section goes through the algorithm in detail. We use a $pp \to \text{dijet}$ sample at $\sqrt{s}=14~\tev$ generated with Pythia 8.176~\cite{Sjostrand:2007gs} to show various distributions. The spectrum is generated such that the $p_T$ of the $2 \to 2$ process is roughly flat across the range $15 - 500~\gev$, in order to maintain reasonable statistics across different kinematic ranges. Pileup events are generated as zero-bias soft QCD events and overlaid onto the hard scatter event.  Further details of the simulation are discussed in \Sec{sec:setup}.

\subsection{The Local Shape}
\label{sec:LocalShape}

For each particle $i$ we define a shape
\begin{eqnarray}
  \label{eq:alpha}
  \alpha_i = \log \sum_{\substack{j\in\ev}} \xi_{ij} \times \,\Theta(R_\text{min}\leq \Delta R_{ij}\leq R_0), \\
  \text{where~} \xi_{ij} = \frac{p_{Tj}}{\Delta R_{ij}}. \nonumber
\end{eqnarray}
Throughout the paper we use $\Theta(R_\text{min}\leq \Delta R_{ij}\leq R_0)$ as a shorthand notation for $\Theta(\Delta R_{ij}-R_\text{min})\times\Theta(R_0-\Delta R_{ij})$, where $\Theta$ is the Heaviside step function.
$\Delta R_{ij}$ is the distance between particles $i$ and $j$ in $\eta\phi$-space and $p_{Tj}$ is the transverse momentum of particle $j$ measured in units of GeV. $R_0$ defines a cone around each particle $i$, so that only particles within the cone enter the calculation of $\alpha_i$. In addition, particles closer to $i$ than $R_\text{min}$ are discarded from the sum, with $R_\text{min}$ effectively serving as a regulator for collinear splittings of particle $i$.  Here we use $R_0=0.3$ and $R_\text{min}=0.02$\footnote{This choice of $R_\text{min}$ is related to typical detector resolutions, as discussed in more detail in \Sec{sec:setup}.}.  
Note that the logarithm is outside of the sum so it plays no role in the infrared-collinear behavior of the variable and just serves to rescale the range. The choice of $\xi_{ij}$ is discussed in more detail in Sec.~\ref{sec:metric}. 

\begin{figure}[thb]
  \centering
  \includegraphics[scale=0.38]{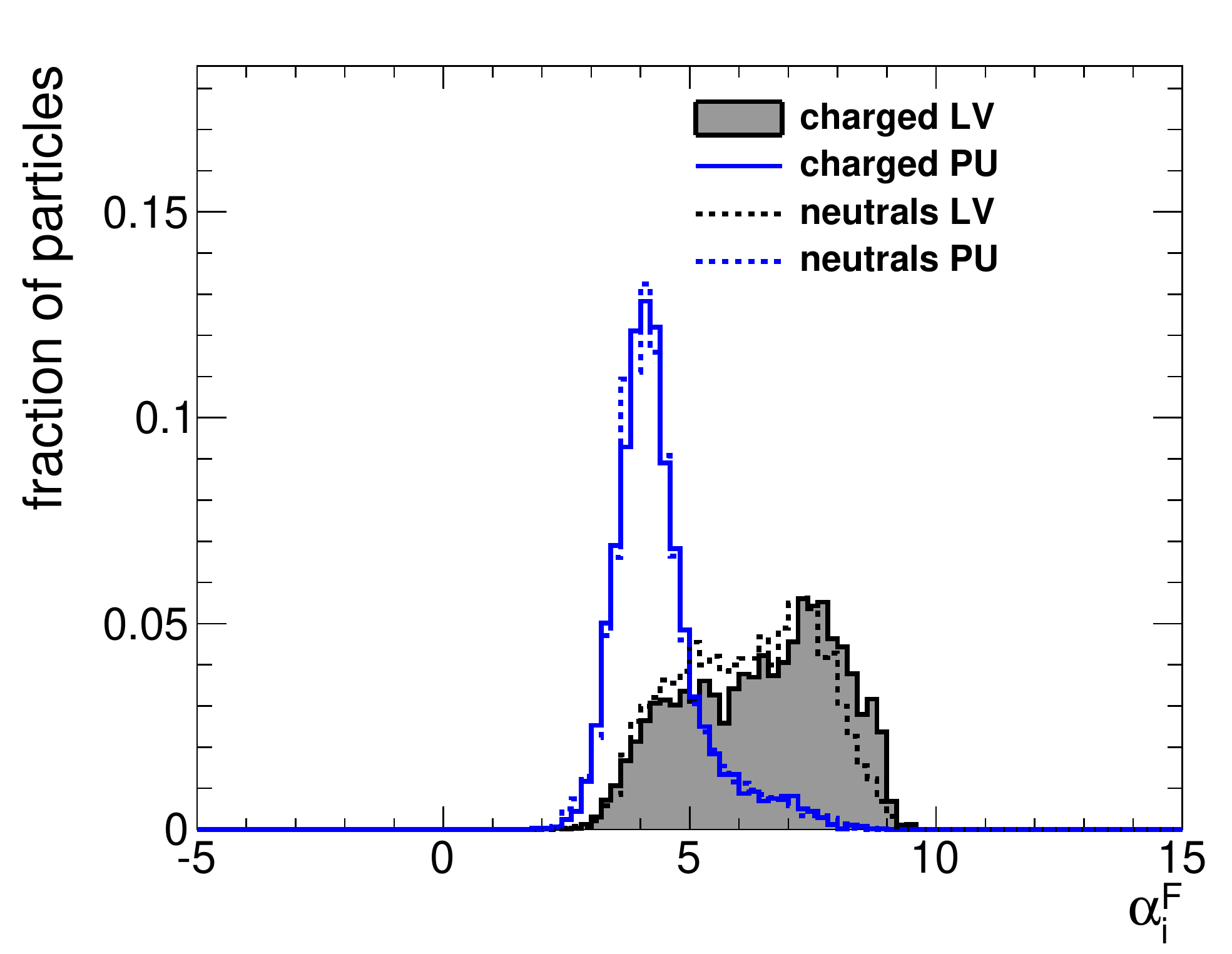}
  \includegraphics[scale=0.38]{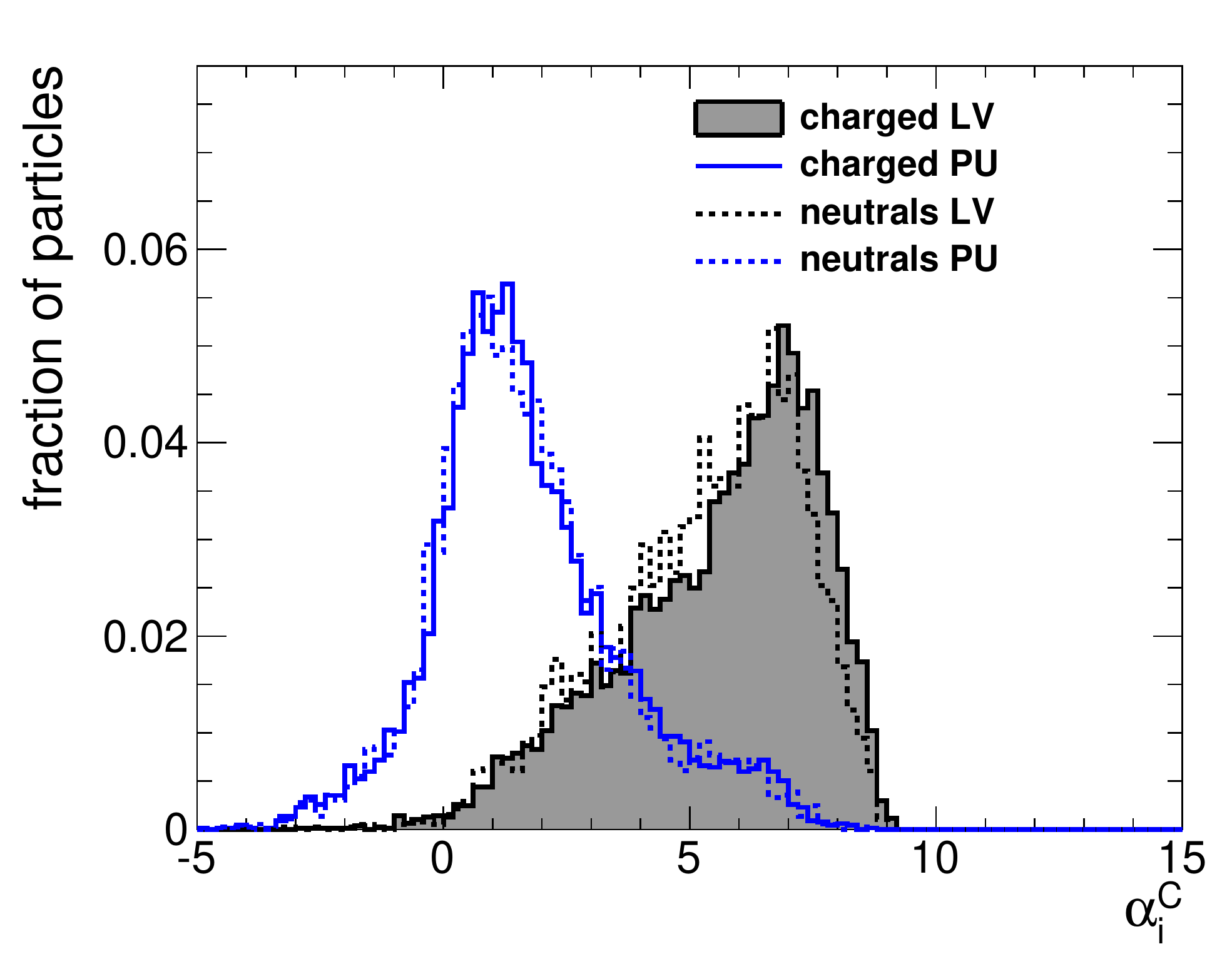}
  \caption{The distribution of $\alpha_i$, over many events, for particles $i$ from the leading vertex (gray filled) and particles from pileup (blue) in a dijet sample.  
  For $\alpha_i^F$ (left) we sum over all particles as defined in \EqsOr{eq:alpha}{eq:alphaFWD}, for $\alpha_i^C$ (right) we sum over charged particles from the leading vertex as defined in \Eq{eq:alphaLV}. Both distributions consider only particles with a $p_T > 1~\gev$. Dotted and solid lines refer to neutral and charged particles respectively.}
  \label{fig:alpha_distr}
\end{figure}

\Fig{fig:alpha_distr} (left) shows a sample distribution of $\alpha$ for particles from the leading vertex and pileup.
%
Due to the collinear singularity of the parton shower, a particle $i$ from a hard physics process is likely to be near other particles from the same parent process so that $\alpha_i$ tends to be larger. On the other hand, we expect pileup particles to have no shower-like structure and to be uncorrelated with particles from the leading vertex and so only to be spatially near by chance\footnote{It is worth noting that stochastic pileup jets can be found by jet algorithms.  This is due to locally high pileup densities rather than a sequence of collinear parton branchings.  As this results in a different radiation pattern on average, pileup jet removal uses differences in jet shapes to remove these pileup jets~\cite{CMS:2013wea}.}. So, $\alpha_i$ tends to be smaller if $i$ is a pileup particle.  In fact, this implies that the ideal version of \Eq{eq:alpha} would sum over particles from the leading vertex and ignore those from pileup.  While we obviously do not know a priori which particles are from the leading vertex, we do have a handle on charged particles in the central ($|\eta| \lesssim 2.5$ for ATLAS and CMS) region of the detector.  In that region, tracking information provides the ability to distinguish charged tracks originating from the leading vertex and charged tracks originating from pileup.  Associating these tracks to particles can be done with the particle flow algorithm~\cite{CMS:2009nxa} which combines measurements from various detector subsystems to define individual candidates\footnote{The use of particles is not strictly necessary.  In principle the algorithm can be performed with calorimeter cells and charged tracks as inputs.  We discuss this later in Sec.~\ref{sec:summary}.}.  Using particle flow, identified candidates can be sorted into three classes: neutral particles, charged hadrons from the leading vertex, and charged hadrons from pileup.  Thus we can use charged particles from the leading vertex as a proxy for all particles from the leading vertex.
To be explicit, in the central region the sum in \Eq{eq:alpha} can be decomposed as
\begin{equation}
  \sum_{j}=\sum_{j\in\text{Ch,PU}}+\sum_{j\in\text{Ch,LV}}+\sum_{j\in\text{Neutral}} ,
  \label{eq:sum}
\end{equation}
where Ch,PU refers to charged pileup, Ch,LV refers to charged particles from the leading vertex, and Neutral refers to all neutral particles both from pileup and the leading vertex.  This leads to defining two versions of $\alpha$ for when tracking information is and is not available.

\begin{eqnarray}
  \label{eq:alphaLV}
  \alpha^{C}_i &=&  \log    \sum_{\substack{j\in\text{Ch,LV}}}\xi_{ij}\,\Theta(R_\text{min}\leq\Delta R_{ij}\leq R_0), \\
  \label{eq:alphaFWD}
  \alpha^{F}_i &=&   \log \sum_{\substack{j\in\ev}}\xi_{ij}\,\Theta(R_\text{min}\leq\Delta R_{ij}\leq R_0).
\end{eqnarray}
Notice that  $\alpha^{F}_i\equiv\alpha_i$ in \Eq{eq:alpha}. Here it is renamed to stress the fact that we use this version of $\alpha_i$ in the forward region of the detector, as opposed to $\alpha_i^C$ which is used in the central region. Effectively, when tracking information is not available, we assume all particles in the sum are from the leading vertex.  While there are noise contributions from pileup, these are suppressed relative to contributions from leading vertex particles by the $p_{Tj}$ in the numerator.  Thus the algorithm can still assign weights in regions where there is no tracking.

\Fig{fig:alpha_distr} (right) shows the distributions of $\alpha^C$. When there are no particles from the leading vertex around particle $i$ to sum over, formally $\alpha_i\to -\infty$. In these cases the particle is assumed to be pileup and discarded from the event\footnote{
The fact that, in practice, the appearance of a single isolated particle occurs much more frequently in pileup (with a moderately low number of pileup interactions) than in hard interactions, supports this assignment scheme.}. The variable $\alpha^C$ has more discrimination power than $\alpha^F$ and is used in the central region of the detector.

There is a second advantage to be gained from tracking information.  For central charged particles, we know the answer as to whether a particle is from pileup or not.  Using only these particles, for a given event we can compute the distribution of both $\alpha^C$ and $\alpha^F$ and then assume that the neutral particles, for which we do not know the answer, belong to a distribution with the same properties.  This assumes the distribution of $\alpha^F$ and $\alpha^C$ is the same for charged and neutral particles, and for central and forward particles.  Neither of these assumptions is exact, but they both can be corrected if necessary. As an example, in \Fig{fig:alpha_distr}, we show the distribution of $\alpha$ for neutral and charged particles separately and we find good agreement overall.

The quantities we use to characterize the distributions on an event-by-event basis are the median and the left-side RMS\footnote{The left-side RMS is computed with $\sum_{\alpha_i<\bar{\alpha}} (\alpha_i - \bar{\alpha})^2$ where $\bar{\alpha}$ is the median of the full distribution.  We use the median and the left-side RMS because these are insensitive to tails in the charged pileup distributions that occur on the right side.  These tails originate from pileup particles inside of jets.}:
\begin{eqnarray}
  \label{eq:alphaSigma_PU1}
    \bar{\alpha}^{F}_\text{PU} &=& \text{median}\{\alpha^{F}_{i\in\text{Ch,PU}}\}, \hspace{4em}
    \sigma^{F}_\text{PU}        =  \text{RMS}\{\alpha^{F}_{i\in\text{Ch,PU}}\}, \\
  \label{eq:alphaSigma_PU2}
    \bar{\alpha}^{C}_\text{PU} &=& \text{median}\{\alpha^{C}_{i\in\text{Ch,PU}}\}, \hspace{4em} 
    \sigma^{C}_\text{PU}        =  \text{RMS}\{\alpha^{C}_{i\in\text{Ch,PU}}\}.
\end{eqnarray}
The characterization of pileup contamination on an event basis is reminiscent of the area subtraction method where average information over an entire event is used to correct individual jets within the event~\cite{Cacciari:2007fd,Cacciari:2008gn}.  In the absence of vertex based discrimination, the median of $\alpha$ can be computed by taking the median over all particles as is done for the area subtraction method.

Because the computation of $\meanf$ and $\rmsf$ is only done for charged pileup, it must be computed in the central region, even though these quantities are used to calculate the weights of particles in the forward region.  Pileup density varies as a function of rapidity and the values of $\meanf$ and $\rmsf$ do not account for this variation.  A proper extrapolation can be performed by estimating the rapidity-dependence in a sample of minimum-bias events.  The weights would then be calculated using the correction $\{ \meanf, \rmsf \} \to f(y_i) \{ \meanf, \rmsf \}$ where $f(y)$ is extracted from minimum-bias data.

\subsection{Particle Weights}
\label{sec:weights}

Having introduced a variable with some separation power between shower-like radiation and pileup-like particles, we will use it to compute a weight for each particle.  The ideal weight is one for leading vertex particles and zero for pileup particles.  Since we are trying to estimate whether a particle is pileup or not given the available information, one can imagine that the weight may not be restricted to one and zero and can be a fractional value. Furthermore, even if one insists on assigning integer weights, in a detector environment neutral particles that are closer than the detector granularity will be treated as a single particle, leading to possible fractional weights.

\begin{figure}[thb]
  \centering
  \includegraphics[scale=0.35]{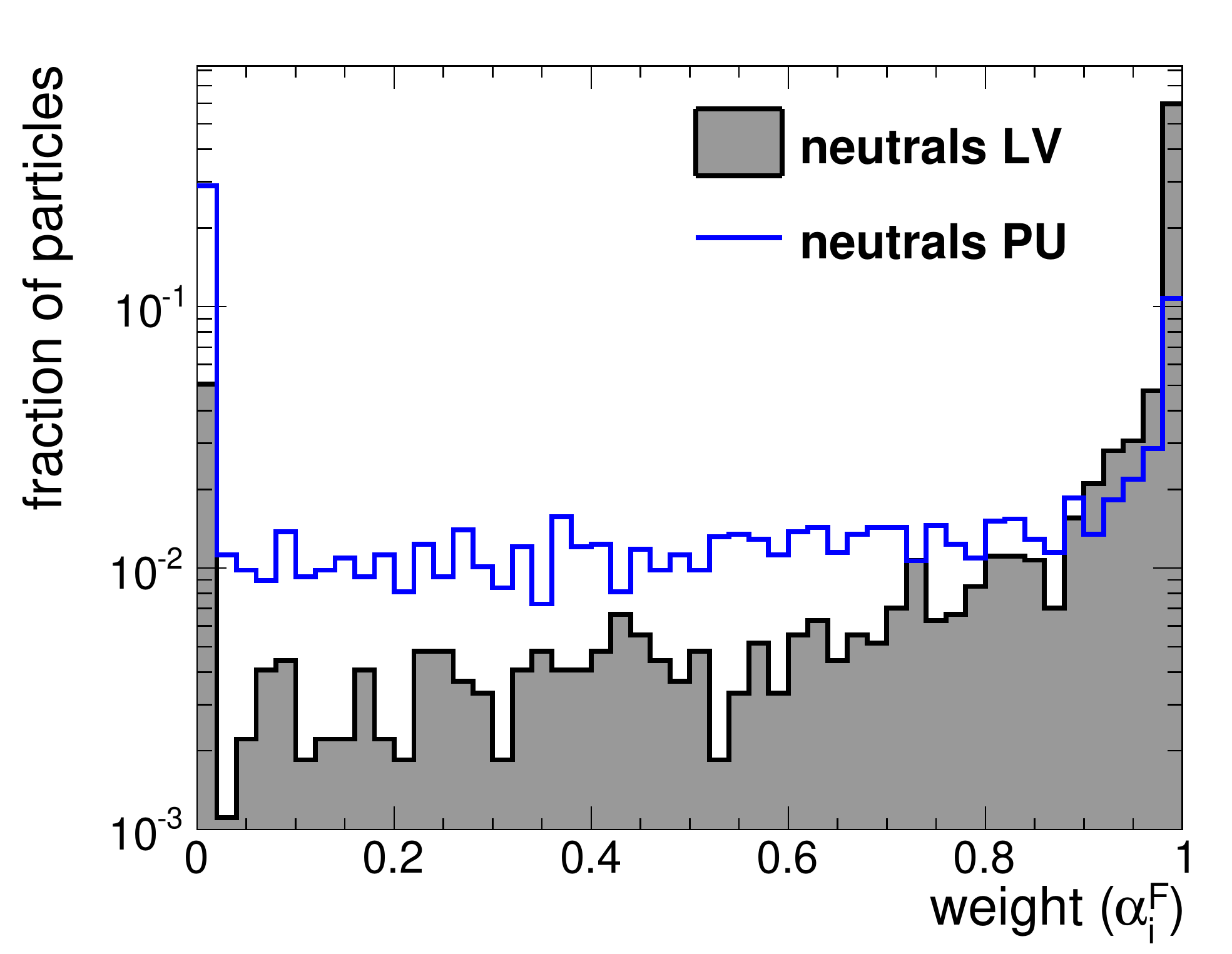}
  \includegraphics[scale=0.35]{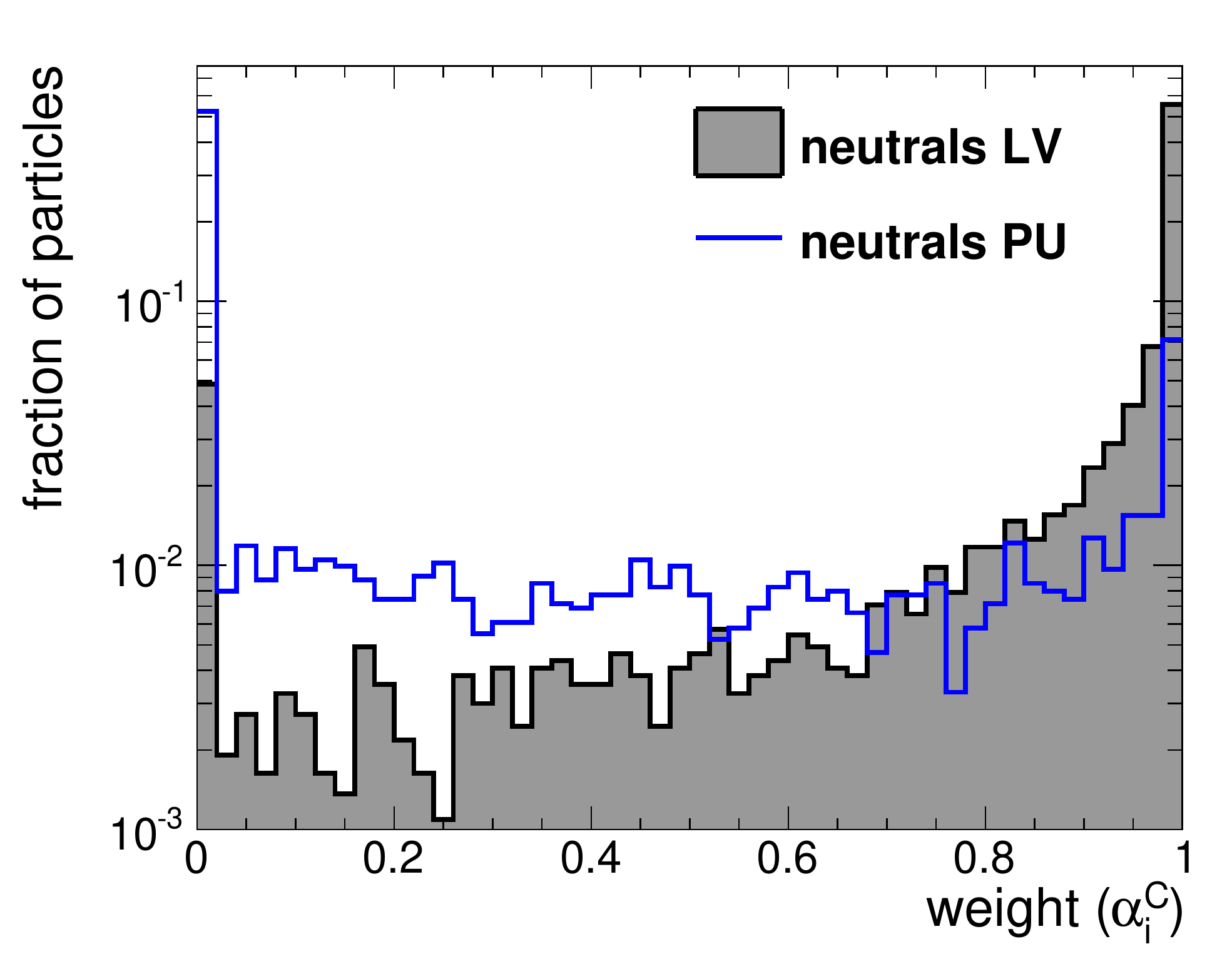}
  \caption{The distribution of weights from \Eq{eq:weight}, over many events, for neutral particles $i$ with $p_T > 1~\gev$ from the leading vertex (gray) and particles from pileup (blue) in a dijet sample. The weights are calculated using $\alpha_i^F$ (left) and  $\alpha_i^C$ (right). In this sample, for weights from $\alpha_i^F$, $30 \%$ ($5 \%$) of neutral PU (LV) particles have $w_i<0.02$ while $10 \%$ ($60 \%$) have $w_i>0.98$.  For weights from $\alpha_i^C$, $50 \%$ ($5 \%$) of neutral PU (LV) particles have $w_i<0.02$ while $5 \%$ ($55 \%$) have $w_i>0.98$.}
  \label{fig:weights}
\end{figure}

In order to define weights, we first introduce the following quantity
\begin{equation}
    \chi_i^2=\Theta(\alpha_i-\bar{\alpha}_\text{PU}) \times \frac{(\alpha_i-\bar{\alpha}_\text{PU})^2}{\sigma_\text{PU}^2},
 \label{eq:chi_2}
\end{equation}
where $\Theta$ is the Heaviside step function.
\Eq{eq:chi_2} measures how far  $\alpha_i$ fluctuates from the pileup median. Fluctuations below the median are considered to be pileup and are assigned a weight equal to zero, as defined below. On the contrary, large fluctuations above the median are very uncharacteristic of pileup and appropriately receive a weight close to $1$. Any intermediate fluctuation above the median is assigned a fractional weight between zero and one. Whenever possible, the $C$ variant of the quantities are used, and everywhere else the $F$ variant is used.  As seen in \Fig{fig:alpha_distr} the distribution of $\alpha$ for pileup looks roughly Gaussian-like.  For this reason \Eq{eq:chi_2} resembles a $\chi^2_{\text{NDF=1}}$ distribution, as the notation suggests.  In fact, interpreting this distribution as a $\chi^2$ distribution lends itself nicely into incorporating additional information, as is discussed in \Sec{sec:addninfo}.  Particles are then assigned a weight given by
\begin{equation}
    w_i = F_{\chi^2,\text{NDF}=1}(\chi_i^2),
    \label{eq:weight}
\end{equation}
where $F_{\chi^2}$ is the cumulative distribution function of the $\chi^2$ distribution. As anticipated, particles with $\chi_i^2=0$ receive a weight $w_i=0$.
\Fig{fig:weights} shows the weight distributions for particles both using $\alpha^F$ (left) and $\alpha^C$ (right).  As expected, the weights are closer to their true value when computed from $\alpha^C$.

At this point we could cut on the weight to decide whether or not to identify a particle as pileup and discard it from the event. In \cite{Krohn:2013lba} it was found that rescaling the particles in subjets was able to correct kinematics and shapes. In light of this, we choose to use the weight in \Eq{eq:weight} to rescale the particle's four-momentum. The complete algorithm proceeds as follows:
\begin{enumerate}
  
  \item The values $\alpha^C_i$ and $\alpha^F_i$ are computed for all charged pileup in the event and the medians and left-side RMS's are computed.

  \item All charged pileup particles are assigned a weight $w_i=0$ and all charged leading vertex particles are assigned a weight $w_i=1$.

  \item The weights of all other particles are calculated using \Eq{eq:weight}.

  \item The four-momentum of each particle is rescaled by its weight $p^\mu_i \to w_i \times p^\mu_i$.

  \item Particles with small weights $w_i < w_{\text{cut}}$ or with low (rescaled) transverse momentum $p_{Ti} < p_{T,\text{cut}}$ are discarded.

  \item The remaining set of rescaled particles is considered the pileup-corrected event.

\end{enumerate}

Let us summarize the parameters of the algorithm. First, we have the cone size $R_0$ which specifies which particles are considered local. Neighboring particles inside a cone are the ones used to calculate $\alpha$. We also have an $R_\text{min}$ cutoff, such that neighboring particles with $\Delta R<R_\text{min}$ are not included in the computation of $\alpha$. In our studies we use $R_0=0.3$ and $R_\text{min}=0.02$. The choice of $R_\text{min}$ is related to typical detector resolutions, as is discussed in \Sec{sec:setup}.
Then we have a weight cut, $w_{\text{cut}}$, below which particles are deemed pileup and a $p_T$ cut, $p_{T,\text{cut}}$. The precise choice of $w_{\text{cut}}$ and $p_{T,\text{cut}}$ depends mildly both on the expected amount of pileup that will be encountered and detector parameters, such as calorimeter granularity.  They can also, in general, be different for the central and forward regions.  In our studies we use $w_{\text{cut}} = 0.1$, $p_{T,\text{cut}} \simeq 0.1 - 1.0~\gev$ (the exact value will be described in \Sec{sec:results}). We have checked that the performance of {\tt PUPPI} algorithm depends weakly on the exact choice of these parameters, with a more significant degradation for much larger values of $R_0$.

One may note that information from the distribution of particles from the leading vertex is primarily ignored.  This is in contrast to matrix-element-like methods like shower deconstruction \cite{Soper:2011cr,Soper:2012pb,Soper:2014rya} which aim to optimize discrimination power by using as much signal and background information as possible.  The specifics of the distributions for leading vertex particles depends on the sample, so we choose not to use the information from the distribution.  In this way, the algorithm is not optimized for any specific signal, but rather looks for general features like a parton shower-like structure, and we expect it to behave consistently across a range of signal topologies.

\begin{figure}
  \centering
  \includegraphics[scale=0.40]{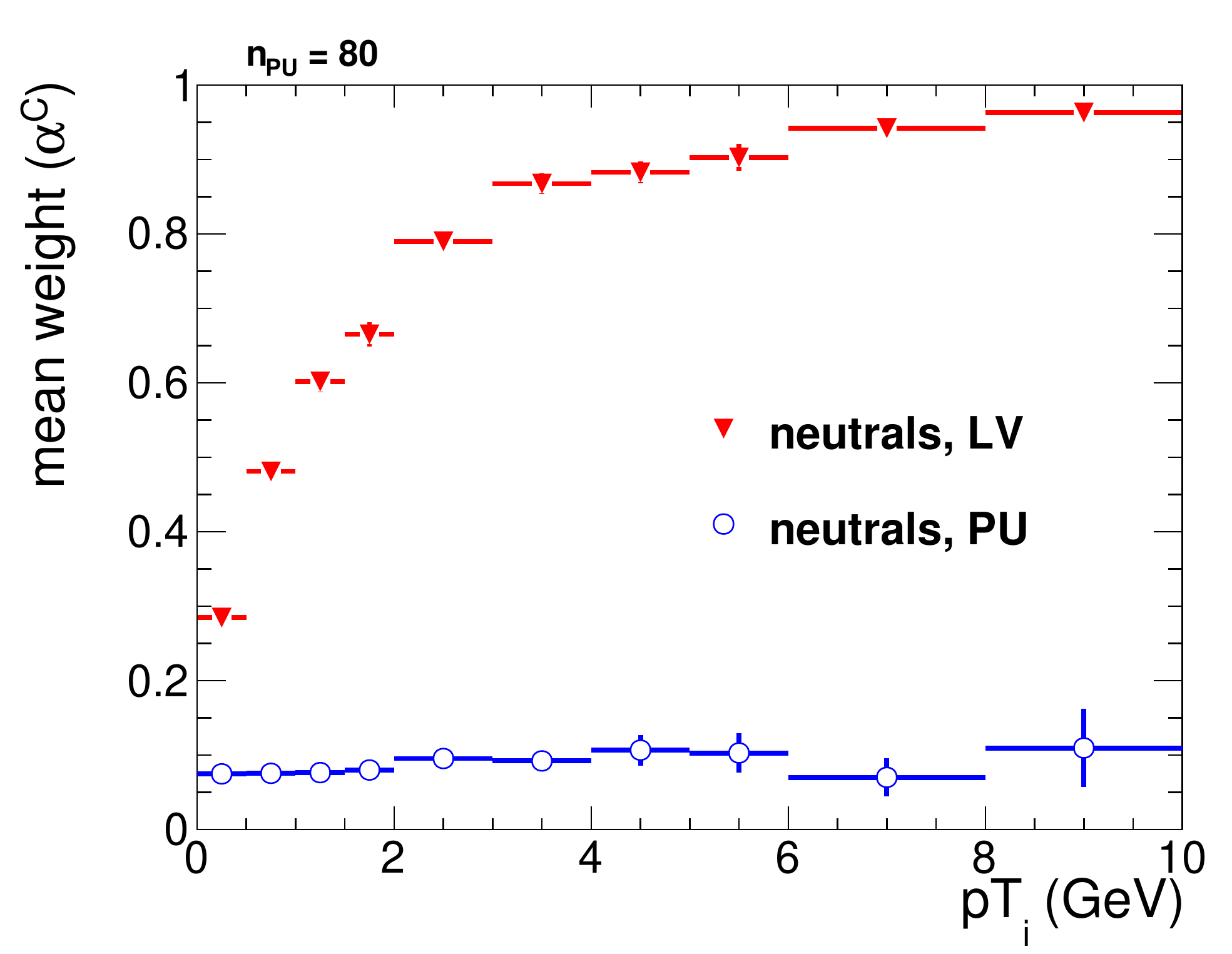}
  \caption{The mean weight, over many events, of neutral particles from the leading vertex (red) and pileup (blue) as a function of the particle's \pt in a dijet sample.}
  \label{fig:weight_vs_pt}
\end{figure}

\subsection{Incorporating Additional Information}
\label{sec:addninfo}

Many pileup removal algorithms are designed assuming a perfect detector and in many cases it is not straightforward to fold in information related to detector efficiencies or limitations.  Using the {\tt PUPPI} algorithm, experimental information can be used to directly modify the weight that is assigned to a particle.  If one interprets the weight as a probability the particle is from the leading vertex (this will be discussed further in \Sec{sec:summary}), then vertex reconstruction efficiencies, for example, may affect this probability.

One advantage to the $\chi^2$ approximation presented above is that it provides a scheme for calculating the weight based on experimental input.  We further make the assumptions that the experimental information is Gaussian-distributed and independent both from the computation of $\alpha$ and other experimental information.  Under these assumptions we can extend the $\chi^2_{\text{NDF}=1}$ approximation to a $\chi^2_{\text{NDF}=N}$ approximation
\begin{equation}
  \chi^{2}_{i,\text{tot}} = \chi^2_i + \sum_{j=2}^N \chi^2_{i,j}
  =
  \chi^{2}_i + \chi^{2}_{i,\text{vertexing}}  + \chi^{2}_{i,\text{calo depth}} + \ldots.
\end{equation}
The weight is then appropriately adjusted to
\begin{equation}
  w_i=F_{\chi^2,\text{NDF}=N}(\chi^2_{i,\text{tot}}).
    \label{eq:weight_n}
\end{equation}
Experimental information that may be useful includes tracking information, calorimeter depth information, and timing information.

\subsection{Choice of Metric}
\label{sec:metric}

In separating pileup from leading vertex particles, it is necessary to identify features that distinguish between them.  Here we consider leading vertex particles to originate from a parton shower.  While the detailed jet structure will depend on the hard process, in the soft and collinear limit the parton shower is universal.  In particular, it includes a soft and collinear singularity leading to the observed collinear structure of jets.  Pileup, on the other hand, contains no hard scale and has no perturbative collinear structure.
This motivates the use of a metric
\begin{equation}
  \alpha_i =\log \sum_{\substack{j\in\ev}} \frac{p_{Tj}}{\Delta R_{ij}^\beta} \times \,\Theta(R_\text{min}\leq\Delta R_{ij}\leq R_0),
  \label{eq:alpha_general}
\end{equation}
where this work uses $\beta=1$. Particles from a parton shower are expected to have a small $\Delta R$ in relation to other particles from the shower, while pileup has no perturbative preference for small $\Delta R$.  The inclusion of $p_{Tj}$ in the numerator is useful for the case where one sums over all particles.  Here, the leading vertex contribution will dominate because the $p_T$ spectrum of pileup falls much faster than leading vertex particles, resulting in the pileup contribution $\alpha$ being supressed by $p_T$.

In \Eq{eq:alpha_general} $\beta$ allows one to tune the relative importance of $p_{Tj}$ vs. $\Delta R_{ij}$.  We have tried many metrics, including those not in the form of \Eq{eq:alpha_general}, and find the one used here to be optimal.

One obvious question that may still arise in the choice of the metric is why $p_{Ti}$ is not used.  After all, we have already stated that the fact the $p_T$ spectrum of pileup falls much faster than the $p_T$ spectrum of leading vertex particles.  Its exclusion from the metric is twofold.  Firstly, it is already used in the algorithm.  After weights are assigned and particles are rescaled, a cut on $p_{Ti}$ is made.  Secondly, we find that one of the reasons we find the weights useful as opposed to just a $p_T$ cut is that the weights tend to not be strongly correlated with $p_T$ in pileup, as shown in \Fig{fig:weight_vs_pt}.  In this way, $\alpha$ uses complementary information to just a $p_{Ti}$ cut. In particular, in trying different metrics, we did try $\alpha_i = p_{Ti}$.  We found its performance to be decent, however it degraded quicker than the $\Delta R$-based metric when calorimeter cell discretization was introduced.

\section{Simulation Details}
\label{sec:setup}

In order to study the performance of our algorithm and compare it to existing methods, we use a sample of $pp \to \text{dijet}$ at $\sqrt{s}=14~\tev$, unless specified otherwise.  Events are generated with Pythia 8.176~\cite{Sjostrand:2007gs}, tune 4C~\cite{Corke:2010yf,Buckley:2011ms}.  The spectrum is generated such that the $p_T$ of the $2 \to 2$ process is roughly flat across the range $15 - 500~\gev$.  This is done in order to maintain reasonable statistics across a range of jet $p_T$ values and to demonstrate the method's utility across different kinematic regimes.  Pileup events are generated as zero-bias soft QCD events using Pythia and overlaid onto the hard scatter event.  The number $n_{\text{PU}}$ specifies the exact number of pileup interactions.  We take as a baseline scenario $n_{\text{PU}} = 80$ pileup interactions overlaid and several results in Sec.~\ref{sec:results} consider this scenario.  We also consider performance versus $n_{\text{PU}}$.  \Secs{sec:kinematics}{sec:shapes} show results on jet kinematics and shapes.  In \Sec{sec:met} we show the algorithm's performance on missing transverse energy ($\MET$).  In this section the sample used is $pp \to Z+\text{jets}$ at $\sqrt{s}=14~\tev$, where the $Z$ decays invisibly to neutrinos.
In order to focus the performance study on pileup mitigation, underlying event is not included in the simulation.

We reconstruct particles in a naive detector simulation.  The detector extends to $|\eta| < 5.0$ and includes a perfect tracker for $|\eta| < 2.5$.  The perfect tracker exactly identifies if a charged hadron is from the leading vertex or from a pileup vertex (in contrast to a real tracker where misidentifications are possible) and has perfect spatial resolution.  Neutral particles are discretized into calorimeter cells of size $0.1 \times 0.1$ in the $\eta\phi$-plane. 

Selecting an appropriate value for $R_{\text{min}}$ is closely tied to the properties of the detector.  The detector itself restricts cells from being closer than approximately $r_{\text{cell}} = 0.1$ from each other.  Similarly, in a real detector the tracking efficiency degrades for distances closer than $r_{\text{track}} \lesssim 0.02$ from each other because for pairs of tracks closer than this distance it becomes possible that one of the tracks is lost.  The distance between cells and tracks, on the other hand, is not necessarily regulated by the detector and could be as small as zero.  Thus $R_{\text{min}}$ directly regulates the cell-track distances and should be chosen as $R_{\text{min}} = \min(r_{\text{track}},r_{\text{cell}})=0.02$ to be consistent with resolutions.  We use $R_{\text{min}}=0.02$ in our simulation for consistency with all inter-object distances and to mock-up the effect of track resolution.

Where particles are clustered into jets, we use Fastjet 3.0.5~\cite{Cacciari:2011ma} and the anti-$k_T$ algorithm~\cite{Cacciari:2008gp} with a radius of $R=0.7$ (AK7).  While smaller jet radii are more common in phenomenological studies, larger jets receive more pileup contamination and are commonplace in substructure studies where correcting more than only a jet's $p_T$ becomes important~\cite{Abdesselam:2010pt,Altheimer:2012mn,Altheimer:2013yza}.  We choose $R=0.7$ as a compromise between these applications.

We define four particle collections from which we can derive algorithmic performance.  They are:
\begin{itemize}

  \item {\tt LV}: Only particles from the leading vertex.

  \item {\tt PFlow}: All particles in the event including those from the leading vertex and pileup.  These are the inputs that would be used in particle flow.

  \item {\tt PFlowCHS}: All particles in the event except for charged particles from pileup (within the tracker volume).  This corresponds to particle flow with charged hadron subtraction.

  \item {\tt PUPPI}: The resulting rescaled particles from the algorithm described in \Sec{sec:algorithm}.
\end{itemize}

The {\tt PFlowCHS} particle collection can be considered the current experimental state-of-the-art.  
We also apply four-vector subtraction to {\tt PFlow} and {\tt PFlowCHS} inputs as will be described in the following section, wherever jet quantities are shown.

\begin{figure}[t]
  \centering
  \includegraphics[scale=0.38]{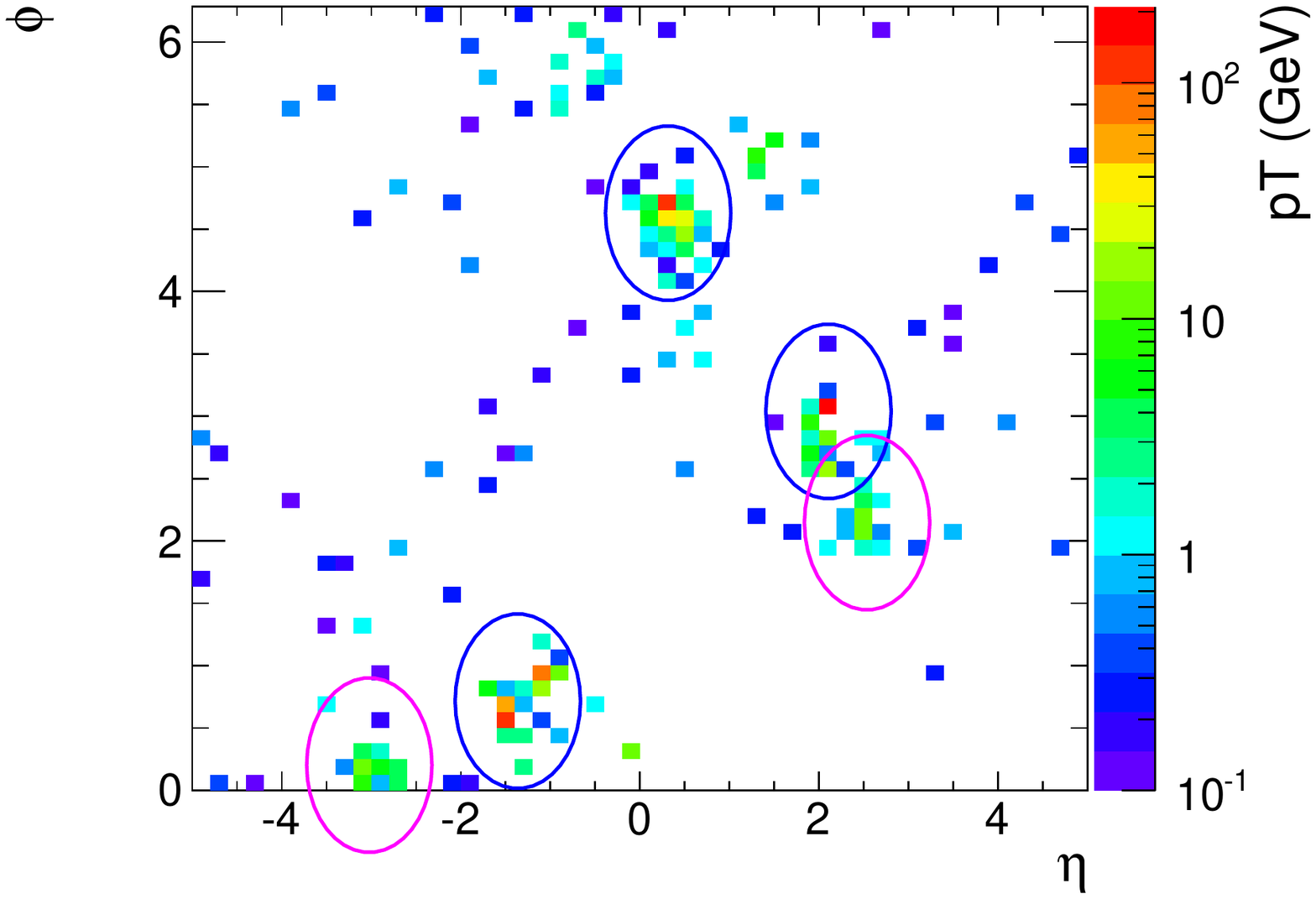}
  \includegraphics[scale=0.38]{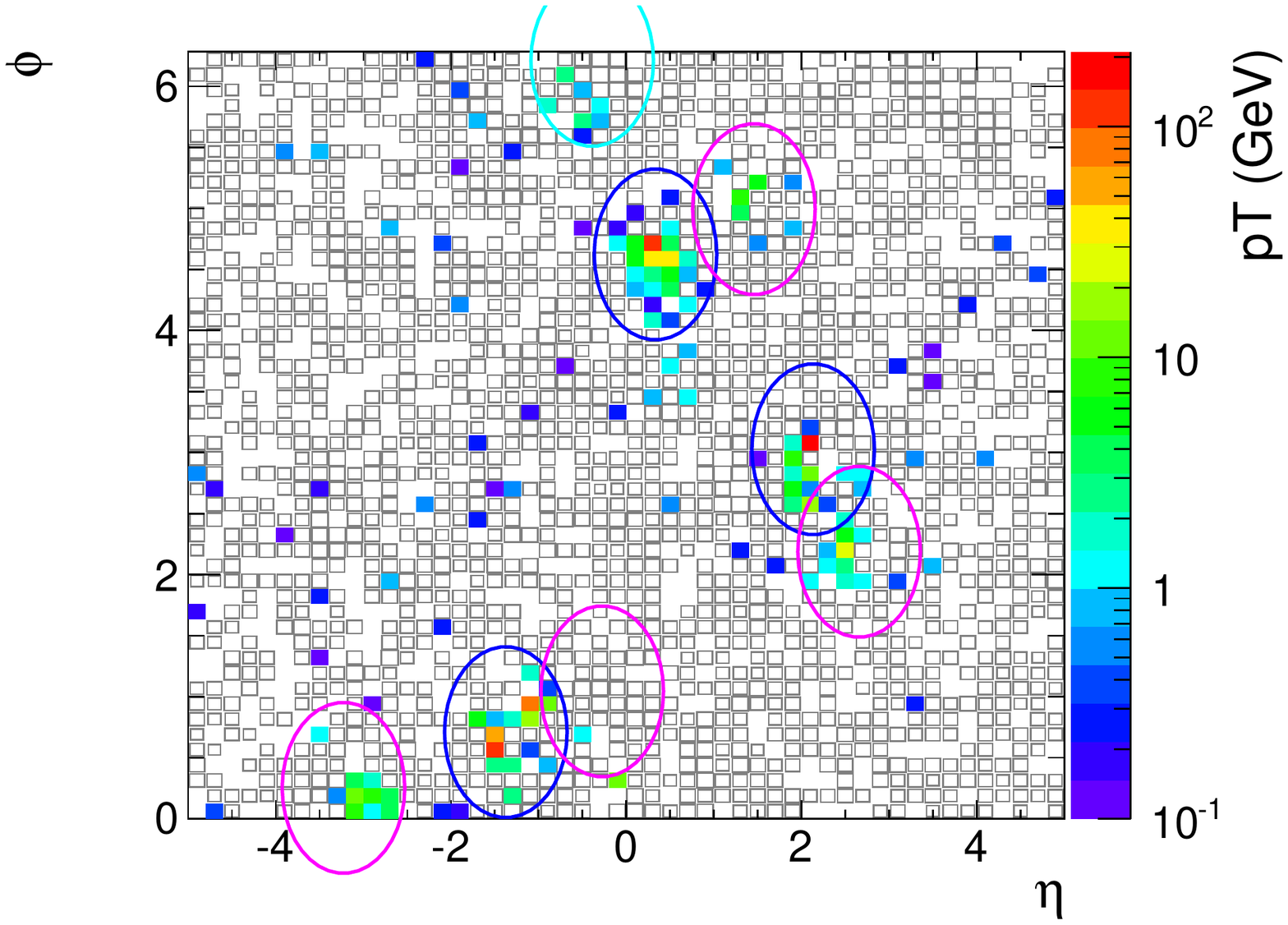}
  \includegraphics[scale=0.38]{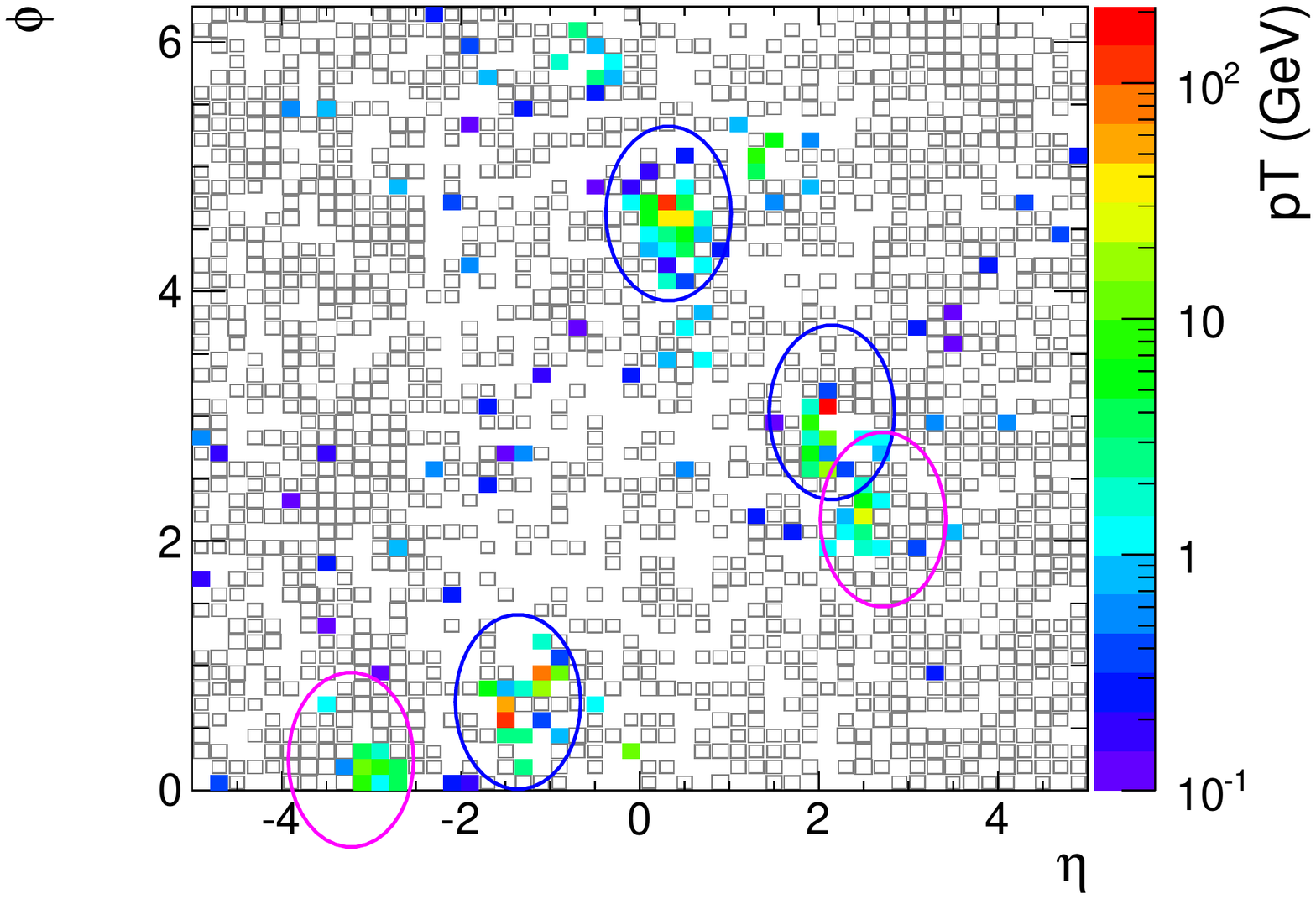}
  \includegraphics[scale=0.38]{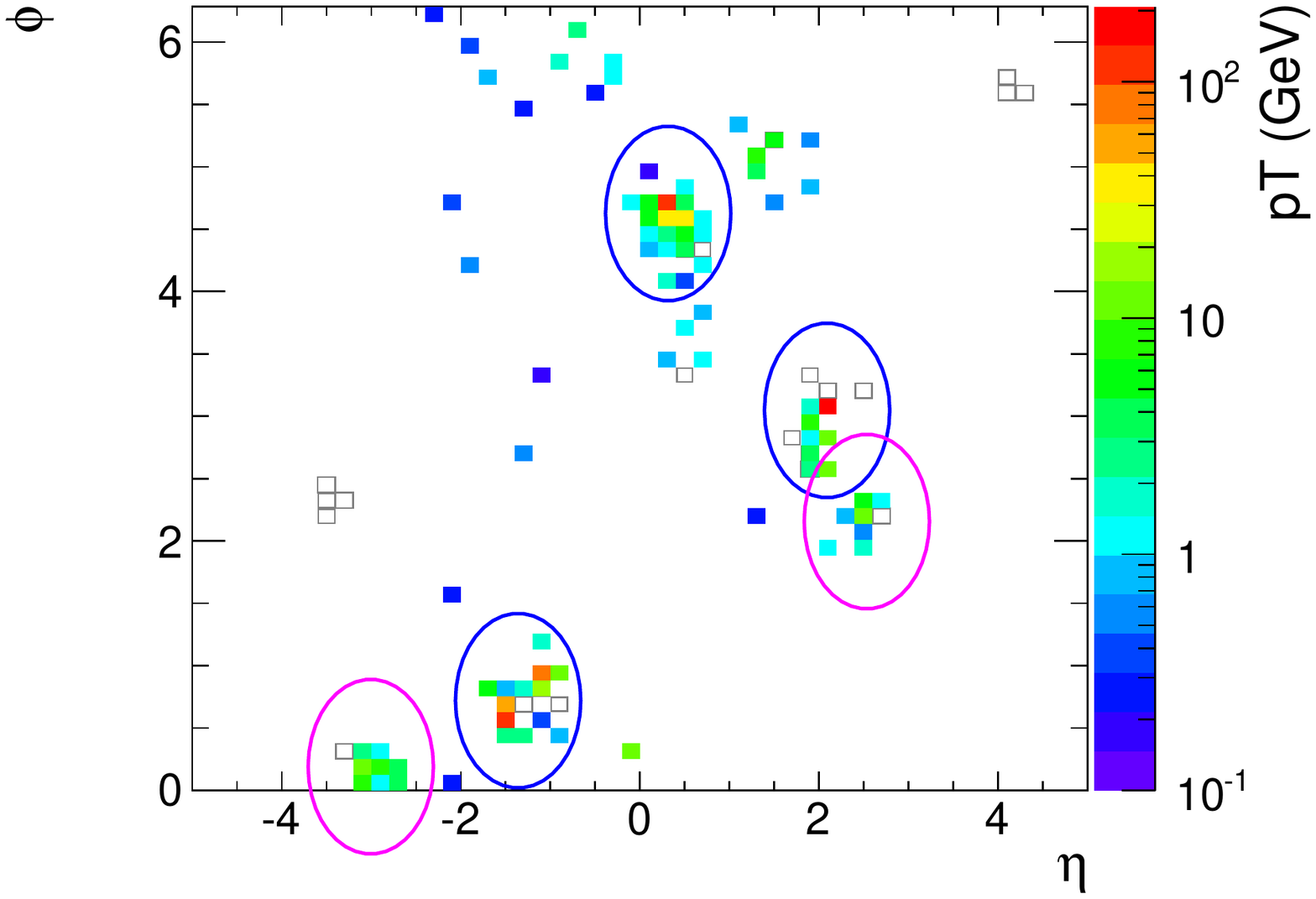}
  \caption{Event display for sample dijet event with $80$ pileup interactions added.  The particle collections shown are {\tt LV} (top left), {\tt PFlow} (top right), {\tt PFlowCHS} (bottom left), and {\tt PUPPI} (bottom right).  Particles from the leading vertex are colored according to their $p_T$, while particles from pileup are uncolored and their size is logarithmically proportional to their $p_T$. The unfilled colored circles show anti-$k_T$ $R=0.7$ jets where the colors denote the $p_T$ bin. The bins $25-50~\gev$, $50-200~\gev$, and $>200~\gev$ correspond to colors of magenta, cyan, and blue respectively.  In the {\tt PFlow} and {\tt PFlowCHS} events, the average value of $p_T$ among the pileup cells is $\sim 0.7~\gev$ and $\sim 0.4~\gev$, respectively.}
  \label{fig:event_display}
\end{figure}

In \Fig{fig:event_display} we show a sample of an event display with $n_{\text{PU}} = 80$ for the four particle collections we consider.  Particles from the leading vertex are drawn with filled squares and colored according to their $p_T$.  Particles from pileup are drawn with unfilled, uncolored squares with their size logarithmically proportional to their $p_T$.  The unfilled colored circles show anti-$k_T$ $R=0.7$ jets where the colors denote the $p_T$ bin.  The bins $25-50~\gev$, $50-200~\gev$, and $>200~\gev$ correspond to colors of magenta, cyan, and blue respectively.

The {\tt LV} plot (top left) shows the original uncontaminated event.  The {\tt PFlow} plot (top right) shows the effect of all pileup particles being added to the event.  The {\tt PFlowCHS} plot (bottom left) shows a reduced pileup density in central region where charged pileup has been removed.  The {\tt PUPPI} plot (bottom right) is an event display that reproduces not only the hard jets from the {\tt LV} collection, but also manages to capture features outside of the jets and remove a large portion of the pileup completely.  The $p_T$ of the jets from {\tt PFlow} and {\tt PFlowCHS} are area subtracted.

\clearpage

\section{Results}
\label{sec:results}

In this section we study the performance of the {\tt PUPPI} algorithm on several jet and event observables.  Where jets are clustered using the {\tt PFlow} collection, they are corrected using the ``safe'' modification of four-vector subtraction \cite{Cacciari:2014jta}\footnote{The results can differ based on the variant of four-vector subtraction used, however, the qualitative conclusions remain unchanged.  In this work we use a modified version of four-vector subtraction presented in \cite{Cacciari:2014jta} which forbids negative masses by setting the mass of (sub)jets to zero in certain cases.}\textsuperscript{,}\footnote{We include corrections due to hadron masses following the method proposed in \cite{Soyez:2012hv}.}.  Subtraction is also applied to jets clustered from the {\tt PFlowCHS}.  In the tracking region for {\tt PFlowCHS}, charged pileup is already removed, so $\rho$ is calculated only from neutral particles.  In the forward region, $\rho$ is computed using all particles.  The jet clustering procedure is run separately on each particle collection.

For {\tt PUPPI} we make the following parameter choices: $R_0 = 0.3$, $R_\text{min}=0.02$,  $w_{\text{cut}} = 0.1$, and choose $p_{T,\text{cut}}$ according to
\begin{eqnarray}
  \text{central: } p_{T,\text{cut}} &=& 0.1~\gev + n_{\text{PU}} \times 0.007~\gev ,\\
  \text{forward: } p_{T,\text{cut}} &=& 0.2~\gev + n_{\text{PU}} \times 0.011~\gev .
\end{eqnarray}
In particular the $p_{T,\text{cut}}$ value has a weak dependence on the amount of pileup in the event and will depend on the granularity and energy resolution of a particular detector.  We tune the values of this cut for our mock detector to minimize the offset between reconstructed observables and {\tt LV} observables (see e.g. missing transverse energy in \Fig{fig:dSumEt}).

\subsection{Jet Kinematics}
\label{sec:kinematics}

\begin{figure}[thb]
  \centering
  \includegraphics[scale=0.38]{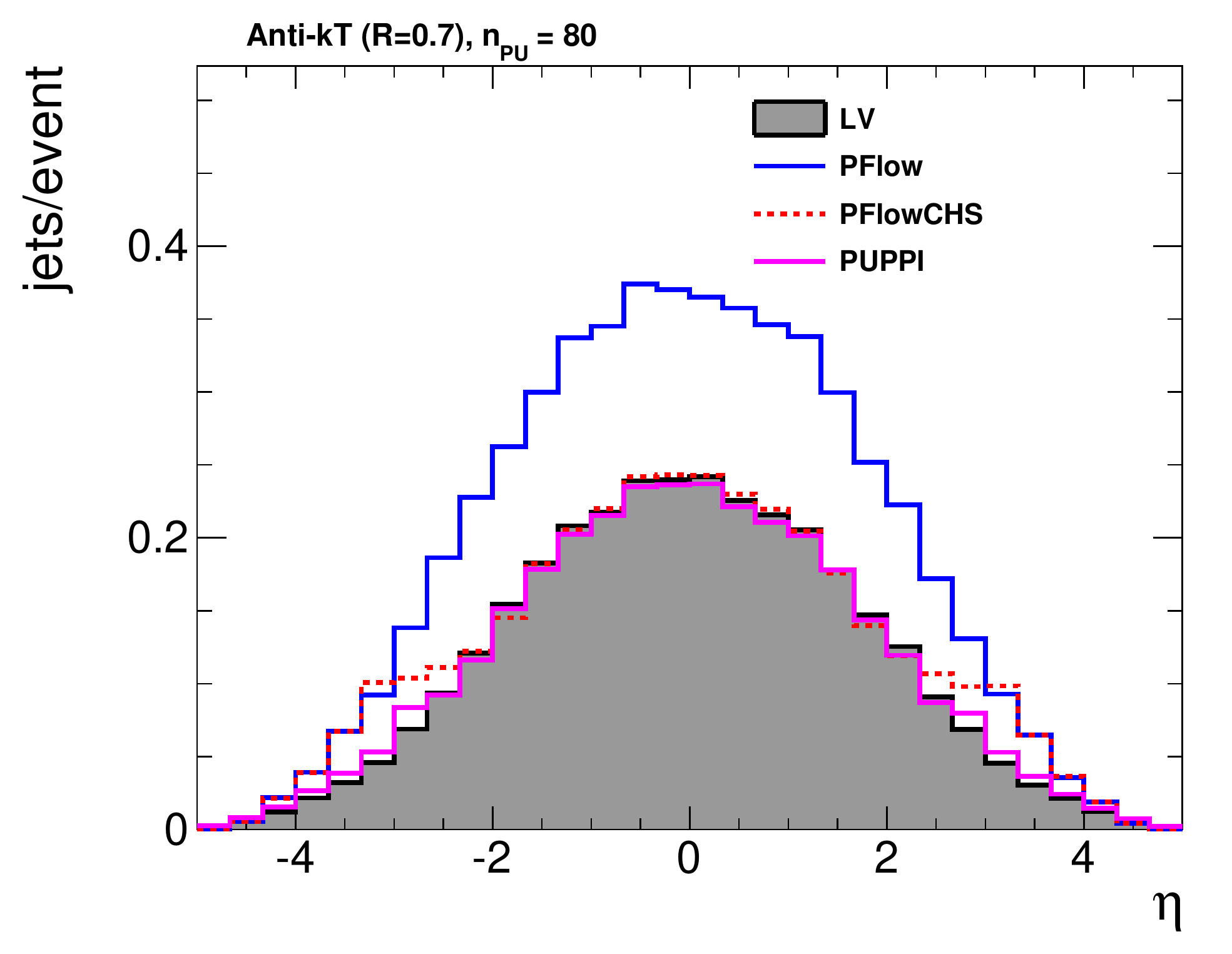}
  \caption{Jet multiplicity as a function of pseudorapidity for $n_{\text{PU}} = 80$.}
  \label{fig:results_eta}
\end{figure}

We start by looking at the jet multiplicity as a function of pseudorapidity shown in \Fig{fig:results_eta} for $n_{\text{PU}} = 80$.  Here all jets with $p_T > 25~\gev$ after the pileup correction techniques are applied are considered.  We see that in pseudorapidity regions where pileup correction is solely from subtraction the jet multiplicity tends to be too high.  This is primarily from high density regions of pileup resulting in pileup jets.  For {\tt PFlow} this occurs across the full rapidity range, while for {\tt PFlowCHS} this only occurs in the forward region where charged hadrons cannot be removed.  The {\tt PUPPI} jet distribution matches the {\tt LV} distribution well across pseudorapidity. 

Next we compare the jet $p_T$ resolution across the methods.  We define the resolution of an observable $\mathcal{O}$ from the particle collection $\mathcal{P}$ to be
%
\begin{equation}
  \text{resolution}(\mathcal{O}_{\mathcal{P}}) =
     \text{RMS} \left\{ \frac{\mathcal{O}_{\mathcal{P}} - \mathcal{O}_{\tt LV}} {\mathcal{O}_{\tt LV}} \right\} . 
  \label{eq:resolution}
\end{equation}
Additionally, in plots where the resolution is cited as fitted $\sigma$, we adopt the common practice of fitting the distribution to a Gaussian and using the standard deviation as the resolution.

\begin{figure}[thb]
  \centering
  \includegraphics[scale=0.38]{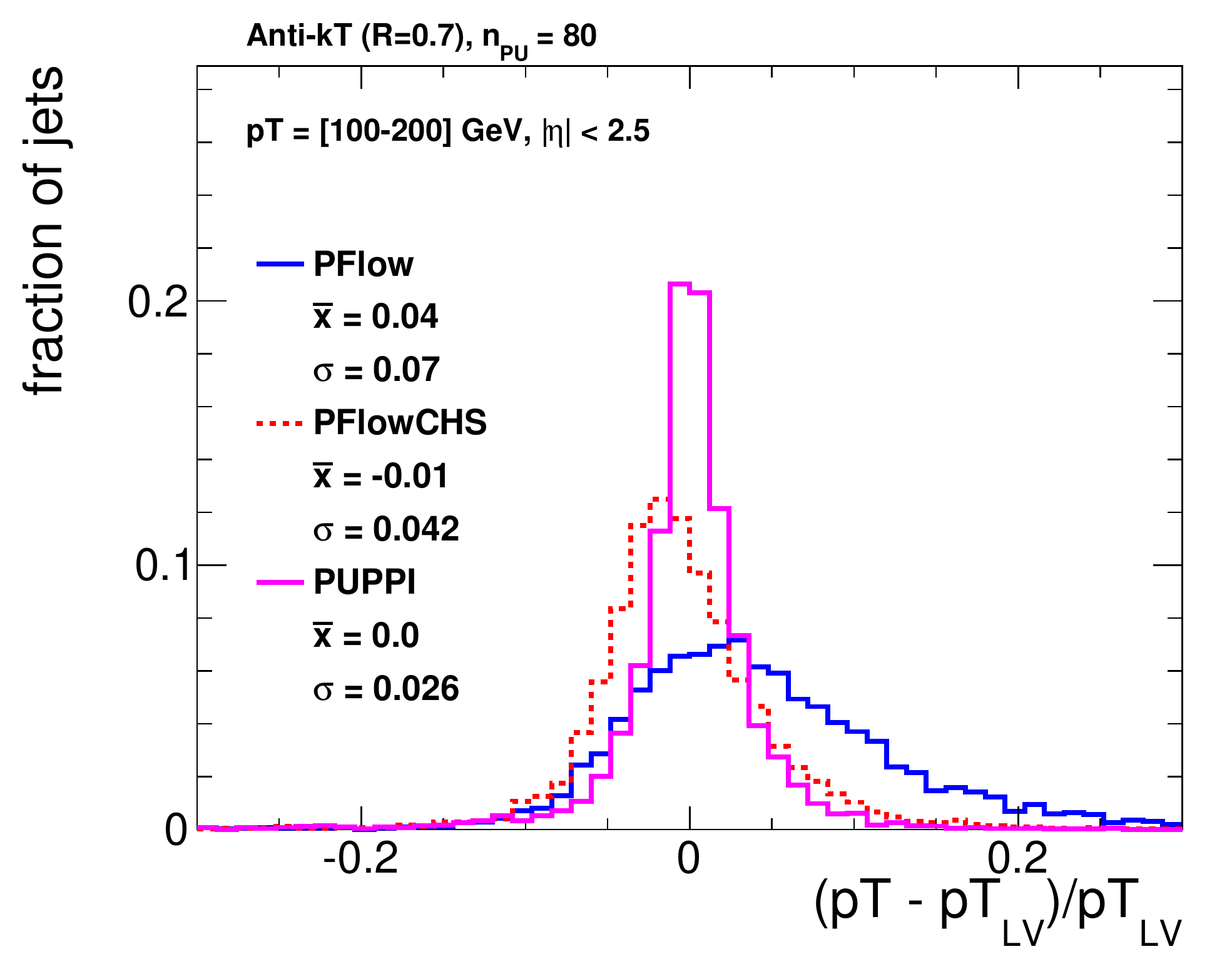}
  \includegraphics[scale=0.38]{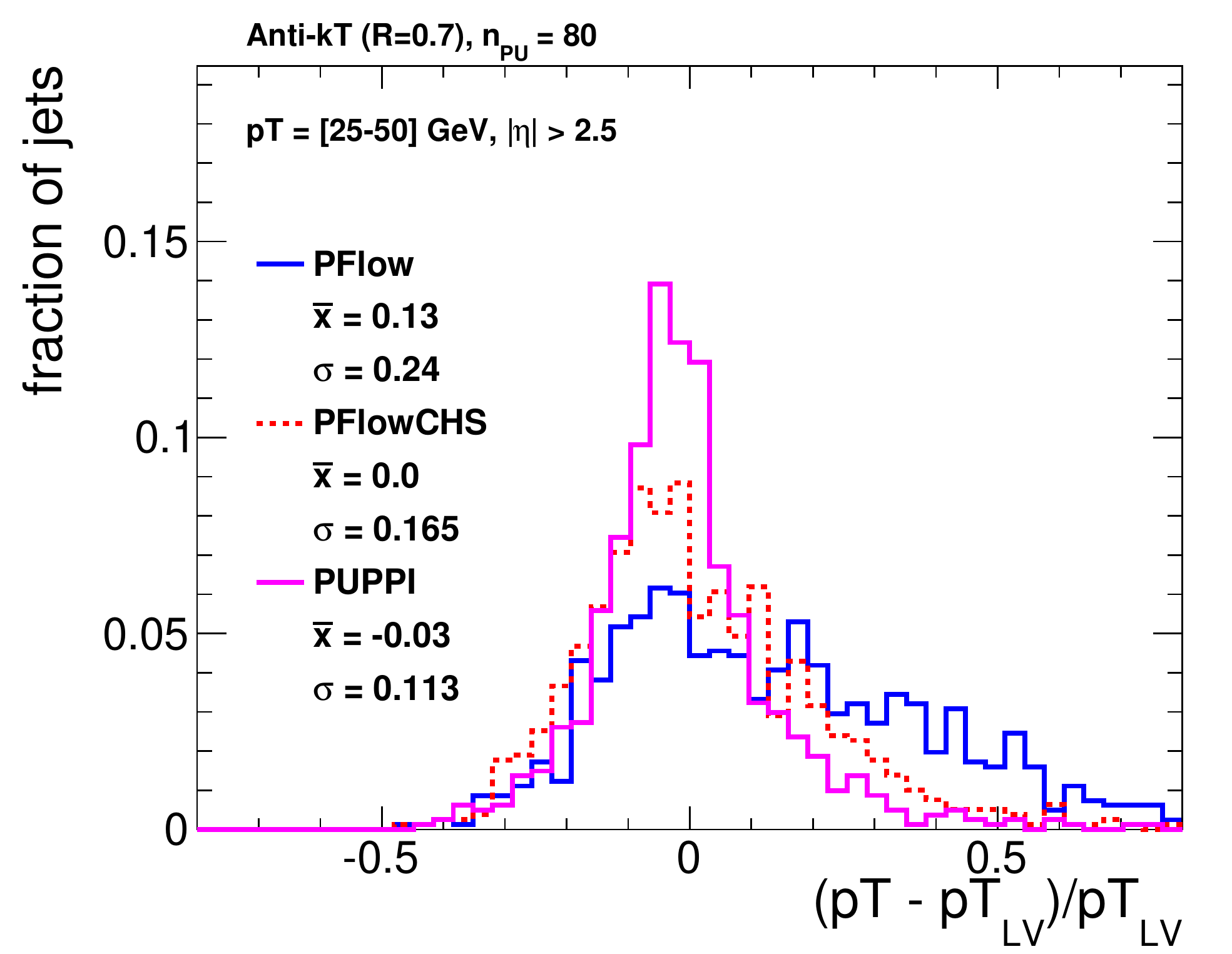}  
  \caption{Jet $p_T$ resolution with $n_{\text{PU}} = 80$ for jets with $100~\gev < p_T < 200~\gev$ and $|\eta| < 2.5$ (left) and jets with $25~\gev < p_T < 50~\gev$ and $|\eta| > 2.5$ (right).}
  \label{fig:results_pt}
\end{figure}

To compare jets from different collections, one needs a scheme to match jets.  We consider jets from two collections matched if they are within $\Delta R = 0.3$ of each other.  \Fig{fig:results_pt} (left) shows the $p_T$ resolution for central jets with $p_T$ between $100$ and $200~\gev$.  The $p_T$ resolution of {\tt PUPPI} is roughly $1.5$ times better than {\tt PFlowCHS} and $2.5$ times better than {\tt PFlow}.  \Fig{fig:results_pt} (right) shows the $p_T$ resolution for forward jets with $p_T$ between $25$ and $50~\gev$.  In both cases the \pt response also tends to be more symmetric than {\tt PFlow} and {\tt PFlowCHS}. Despite the fact that there is no tracking information in the forward region, the {\tt PUPPI} algorithm is able to maintain an improvement over subtraction even in the forward region. We also note that the improvement in {\tt PFlowCHS} over {\tt PFlow} in the forward region is due to the partial tracking information that {\tt PFlowCHS} jets have near the tracker boundary.
%

\begin{figure}[thb]
  \centering
  \includegraphics[scale=0.38]{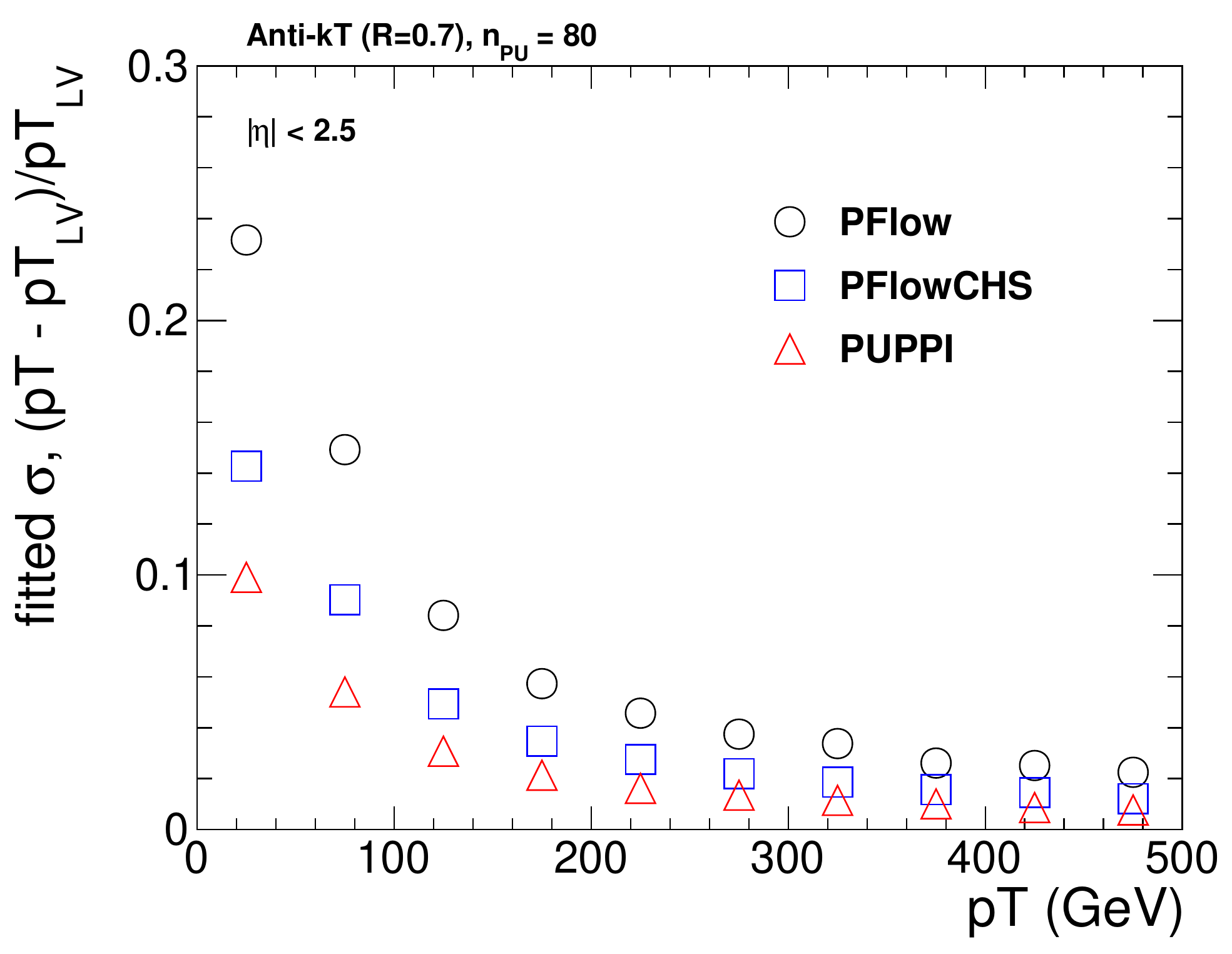}
  \includegraphics[scale=0.38]{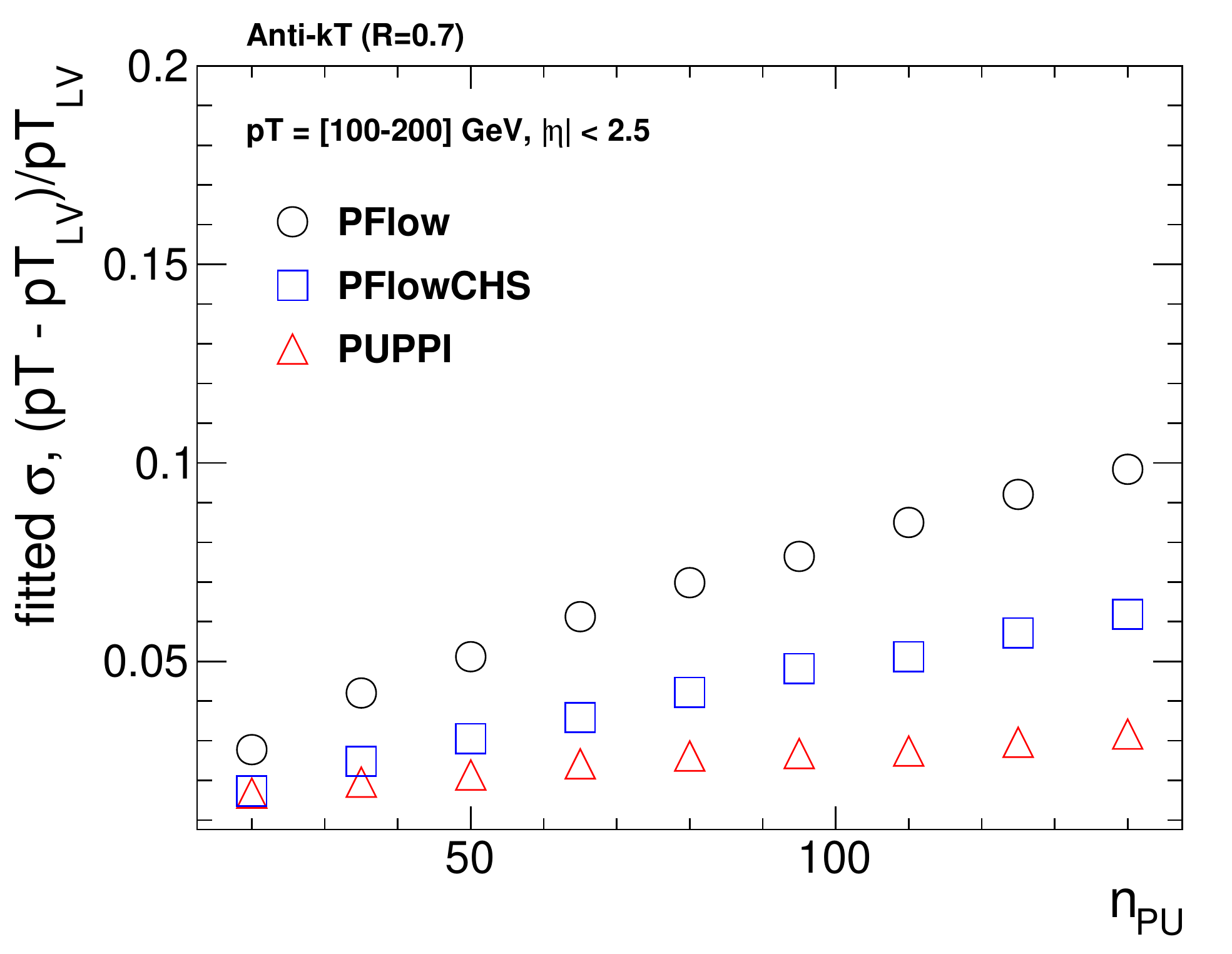}
  \caption{Jet $p_T$ resolution vs. $p_T$ (left) for $n_{\text{PU}}=80$ for $|\eta|<2.5$ and jet $p_T$ resolution vs. number of pileup interactions (right) for jets with $p_T$ between $100$ and $200~\gev$ for $|\eta|<2.5$.}
  \label{fig:results_ptres}
\end{figure}

Next we show the $p_T$ resolution as a function of $p_T$ for central jets in \Fig{fig:results_ptres} (left).  We show that the improvements found above hold across a wide kinematic range.  In \Fig{fig:results_ptres} (right) we show the $p_T$ resolution as a function of number of pileup interactions.  For low levels of pileup we see that the {\tt PUPPI} algorithm does not offer much of an improvment over existing methods.  This is for two reasons.  Firstly, at low levels of pileup there is not much improvement to make.  Secondly, in low pileup environments, there is less information available locally just due to the lack of pileup.  This means the $\alpha$ distribution is not as well populated and the uncertainty on $\sigma_{\text{PU}}$ is larger.

\subsection{Jet Shapes}
\label{sec:shapes}

Similar to our study of $p_T$ distributions, we can study resolution and its pileup dependence for jet shapes.  Here we show results for jet mass which is considered a reasonable proxy for generic jet shapes and is used in many applications such as boosted object tagging (see \cite{Abdesselam:2010pt,Altheimer:2012mn,Altheimer:2013yza} and references therein).

\begin{figure}[thb]
  \centering
  \includegraphics[scale=0.35]{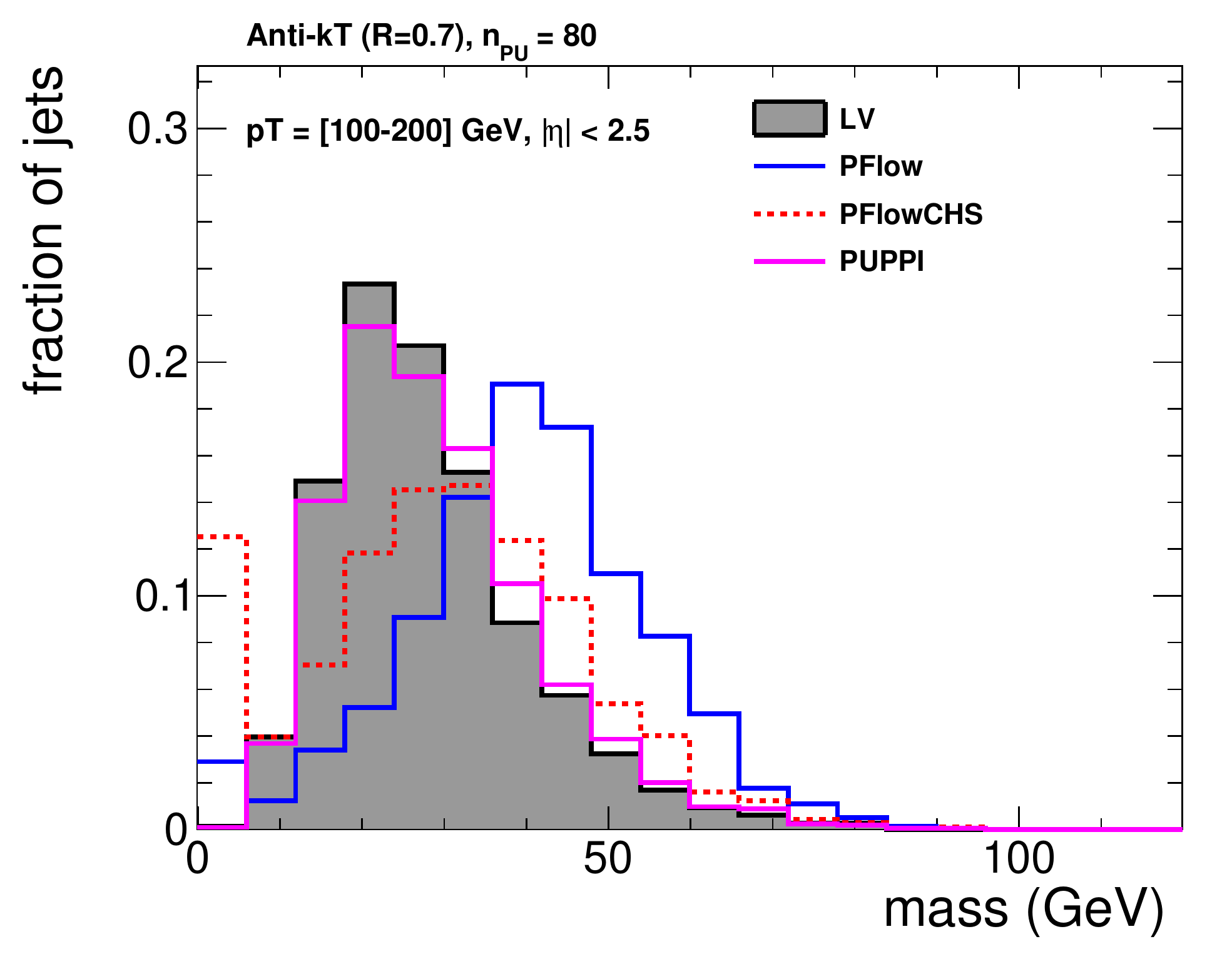}
  \includegraphics[scale=0.35]{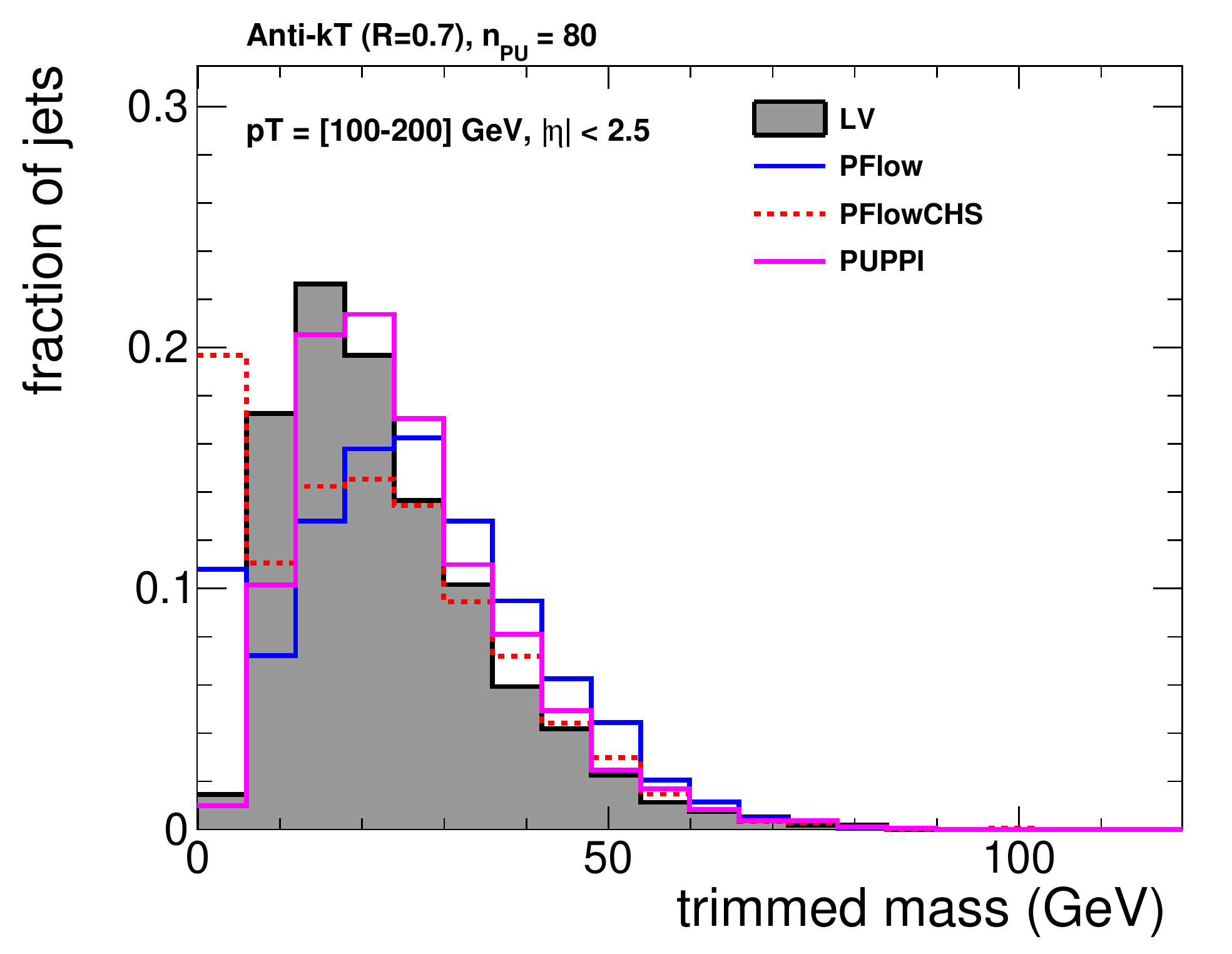}
  \caption{The single jet mass resolution for $n_{\text{PU}} = 80$ for jets with $100~\gev < p_T < 200~\gev$ and $|\eta| < 2.5$ for plain jet mass (left) and trimmed jet mass (right).}
  \label{fig:results_m}
\end{figure}

First we look at jet mass for central jets with $100~\gev < p_T < 200~\gev$.  The distribution is shown in \Fig{fig:results_m} (left).  Here we see that {\tt PUPPI} is not only able to correct the mean of the distribution, but also the distribution itself.  \Fig{fig:results_m} (right) shows the results of {\tt PUPPI} on trimmed mass.  Trimming is performed on jets from all collections, including {\tt LV}, using $r_{\text{sub}} = 0.2$ and $f_{\text{cut}} = 0.05$.  For jets from {\tt PFlow} and {\tt PFlowCHS} subtraction is applied to the trimmed jet.  Even with the application of grooming, {\tt PUPPI} distributions do a consistent job of restoring distributions near to their {\tt LV} distributions.  We regard this as a positive indication that {\tt PUPPI} is returning a consistent event interpretation.  

\begin{figure}[thb]
  \centering
  \includegraphics[scale=0.35]{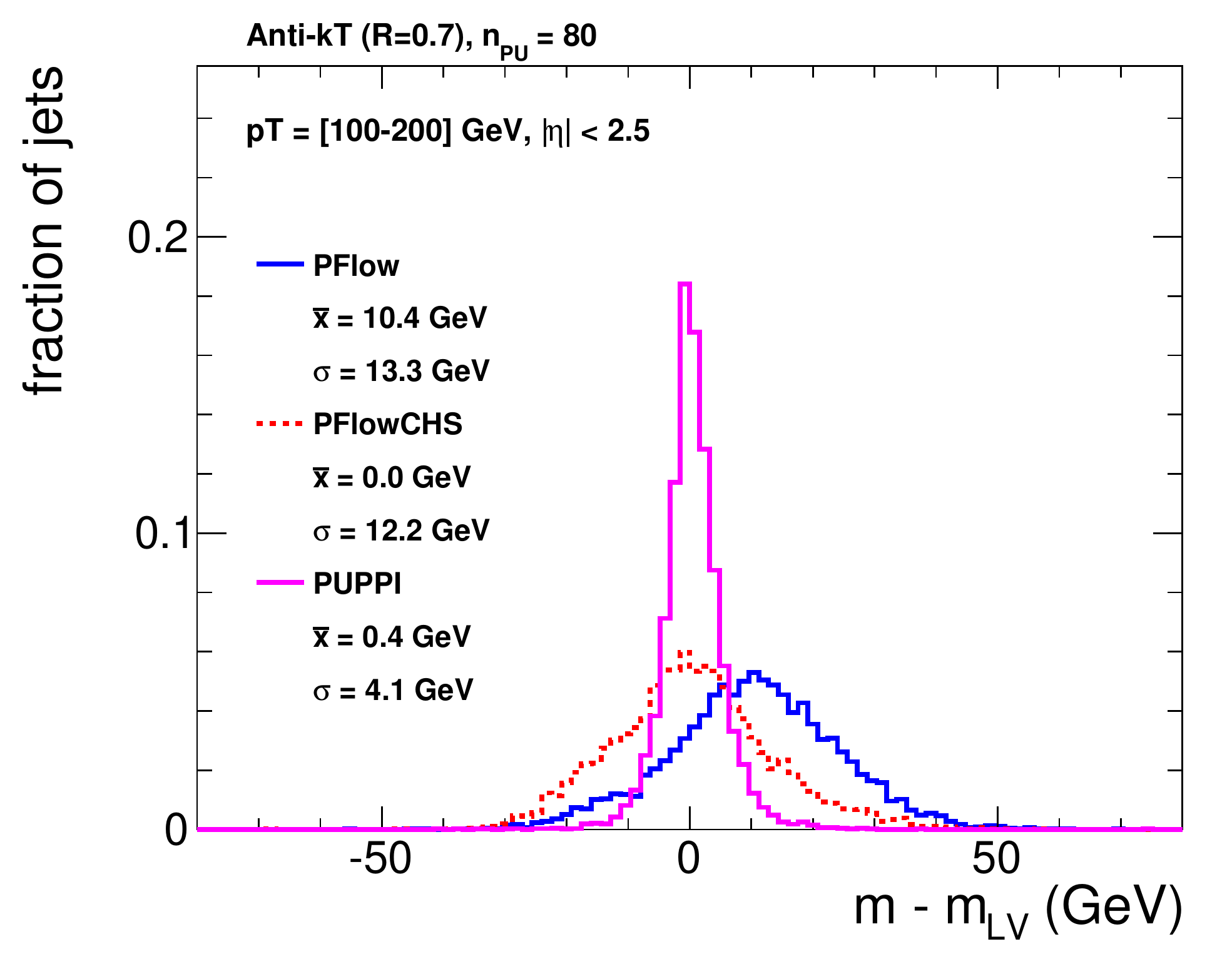}
  \includegraphics[scale=0.35]{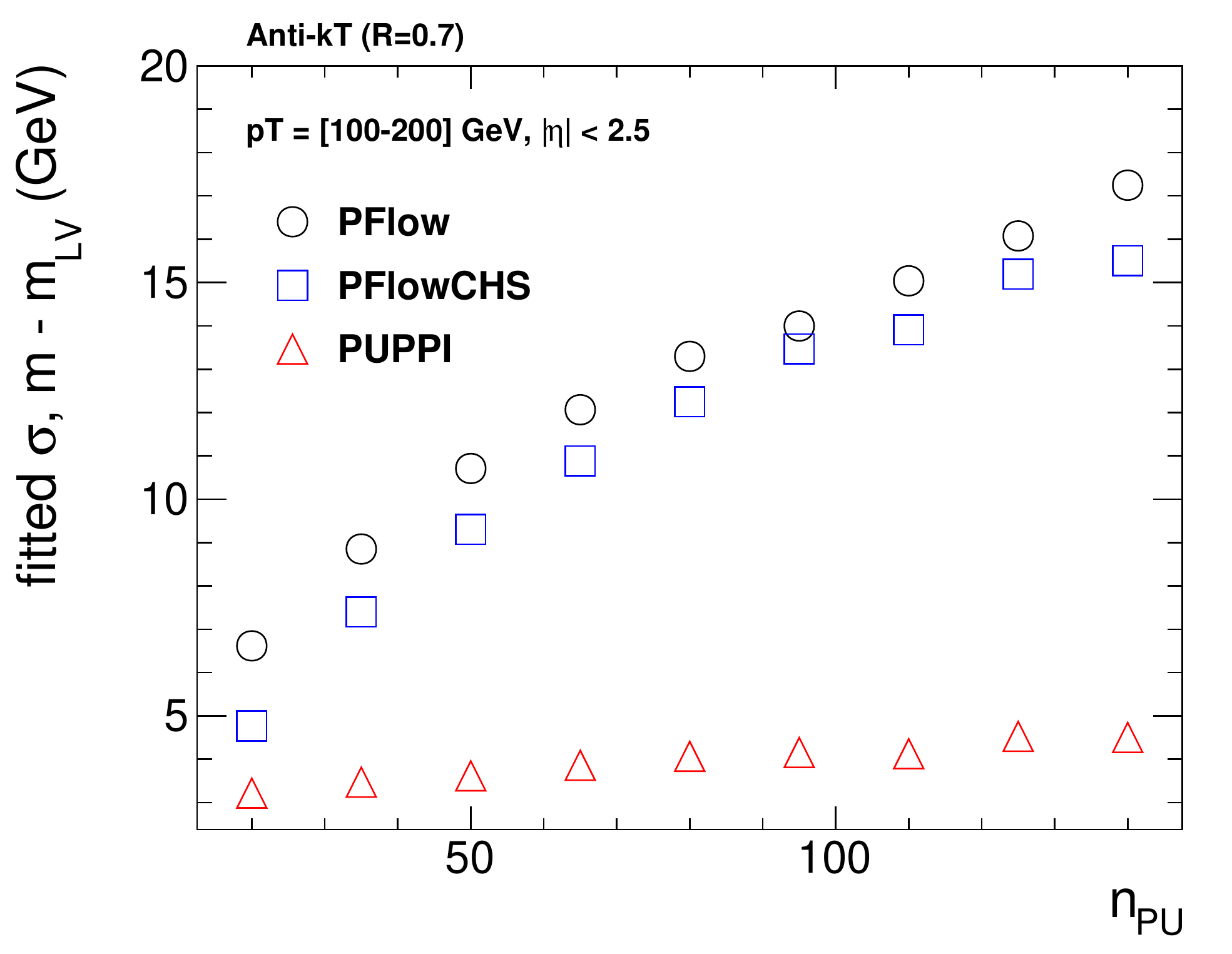}
  \caption{The single jet mass resolution for $n_{\text{PU}} = 80$, for jets with $100~\gev < p_T < 200~\gev$ and $|\eta| < 2.5$ (left) and jet mass vs. number of pileup interactions for jets with mass between $100$ and $200~\gev$.}
  \label{fig:results_mres}
\end{figure}

In \Fig{fig:results_mres} (left) we show the mass resolution\footnote{For mass and missing transverse energy resolutions we use $\text{resolution}(\mathcal{O}_{\mathcal{P}}) = \text{RMS} \{ \mathcal{O}_{\mathcal{P}} - \mathcal{O}_{\tt LV} \}$ as opposed to \Eq{eq:resolution} to avoid divergent behavior as $\mathcal{O}_{\tt LV} \to 0$.} for jets with $p_T$ between $100~\gev$ and $200~\gev$ at $n_{\text{PU}}=80$.  We find that the {\tt PUPPI} jet mass resolution is improved with respect to the other inputs. 
\Fig{fig:results_mres} (right) plots the mass resolution as a function of number of pileup interactions where the mass resolution from {\tt PUPPI} is relatively stable as a function of $n_{\text{PU}}$.

\subsection{Missing Transverse Energy}
\label{sec:met}

We now look at an event quantity, the missing transverse energy ($\MET$), which is interesting from both a theoretical and an experimental point of view. From the theoretical perspective, missing transverse energy is one of the main signatures of new physics.  For example, in $R$-parity conserving supersymmetry, every event in which superpartners are pair-produced the two lightest supersymmetric particles in the final state appear as missing transverse energy.  Additionally for standard model measurements, $\MET$ plays a role in many analyses, such as the $W$ mass measurement~\cite{Aaltonen:2013iut}, the Higgs to $WW$ discovery~\cite{Chatrchyan:2013iaa,ATLAS-CONF-2013-030} and the Higgs to $\tau\tau$ evidence~\cite{ATLAS-CONF-2013-108,Chatrchyan:2014nva}. On the experimental side, $\MET$ is challenging because it compounds errors from the measurement of all objects in the event, both pileup and non-pileup alike. In the presence of pileup, the $\MET$ resolution rapidly degrades because the full energy of the additional pileup events is incorporated into the event~\cite{CMS-PAS-JME-12-002,ATLAS-CONF-2014-019}. 

Attempts at reducing the impact of pileup on the $\MET$ resolution are typically more difficult than on jets, because traditional approaches that work on jets breakdown.  The pileup component of events has a natural tendency to have near zero $\MET$.  Applying a method that reconstructs $\MET$ from a fraction of the particles in the event, {\it e.g.} charged hadron subtraction, breaks the cancellation between neutral and charged pileup particles resulting in large distortions in $\MET$ measurements.
In order to mitigate the effects of pileup in $\MET$, both ATLAS and CMS have resorted to approaches that rely on combinations of various methods of calculating $\MET$~\cite{CMS-PAS-JME-12-002,ATLAS-CONF-2014-019,Chatrchyan:2012ty}.  Such methods, either through a linear combination of different $\MET$ variables or through a boosted decision tree regression, can lead to a reduction of the pileup dependence on the $\MET$ resolution by a factor of roughly three. These calculations are typically quite elaborate and rely on the commissioning of $10 - 20$ additional $\MET$ related variables. 

To compare the performance of $\MET$ observables, we use a sample of events with large hadronic recoil and well-defined $\MET$, in this case $pp \to Z j$ where the $Z$ decays to neutrinos and has transverse momentum in the center of mass frame $p_T(Z)>350~\gev$.  The missing transverse energy is constructed from negative vector sum of the particle transverse momenta
\begin{equation}
  \vec{E}_T^{\text{miss}} = -\sum_{i\in\ev} \vec{p}_{Ti},
  \label{eq:met}
\end{equation}
where the length of this vector is denoted $\MET = |\vec{E}_T^{\text{miss}}|$.  Another related variable is the scalar sum of transverse energies
\begin{equation}
  \sum E_T = \sum_{i\in\ev} |\vec{p}_{Ti}|.
  \label{eq:sumet}
\end{equation}

\begin{figure}[thb]
  \centering
  \includegraphics[scale=0.38]{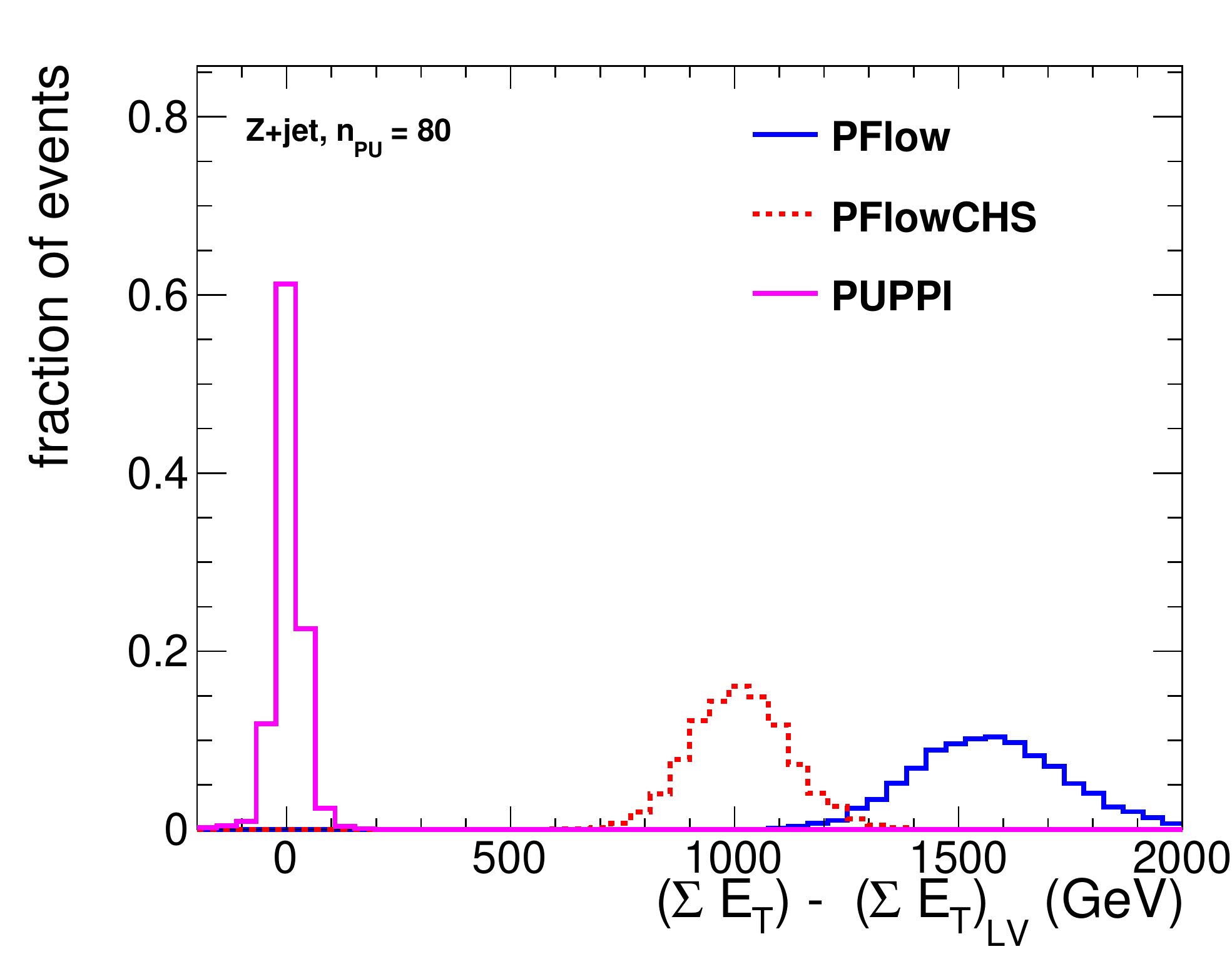}
  \includegraphics[scale=0.38]{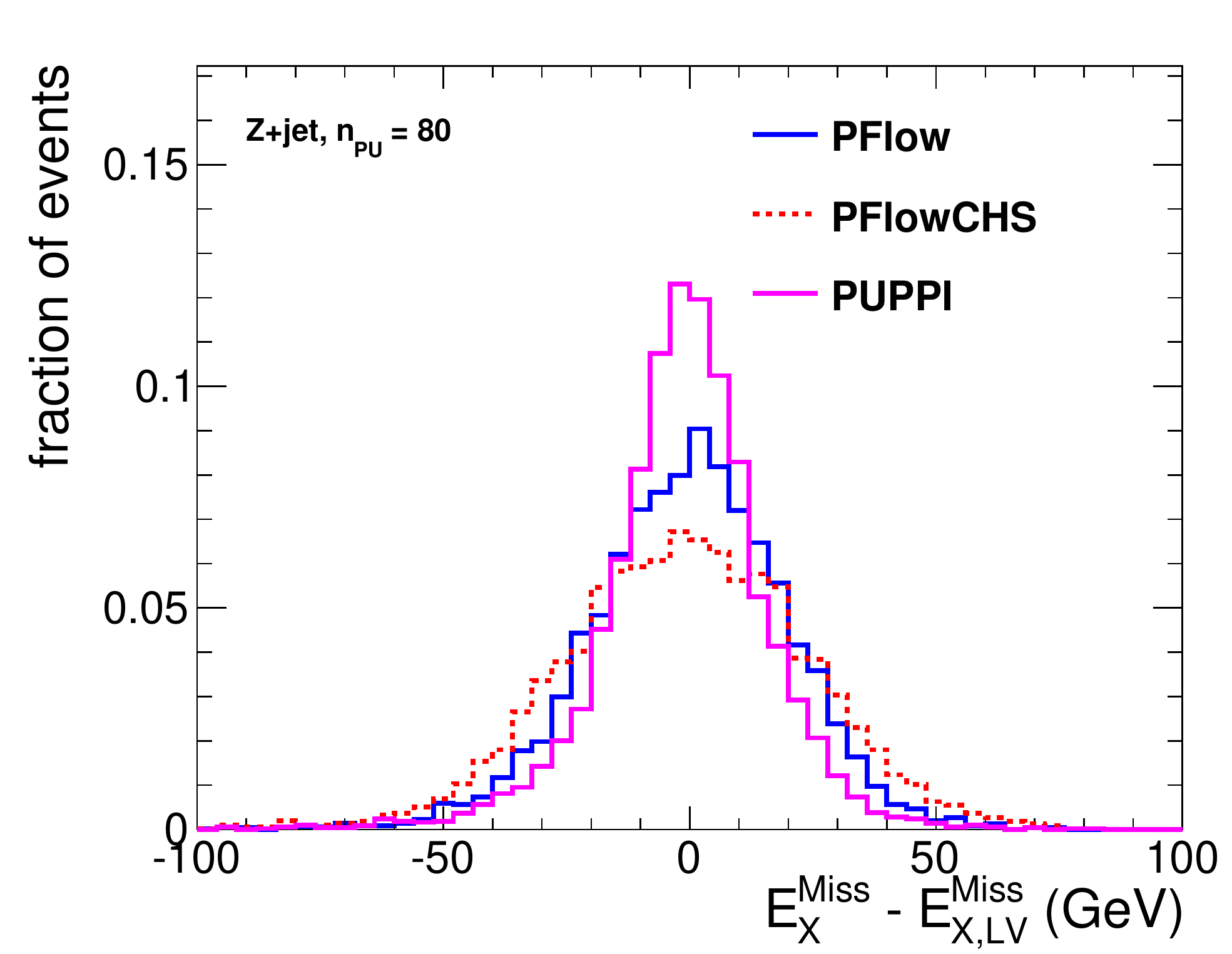}  
  \caption{The resolution of $\sum E_T$ (left) and the resolution of $\METx$ (right) in $Z+\text{jets}$ events with $n_{\text PU}$ = 80.}
  \label{fig:dSumEt}
\end{figure}

We show the resolution in \Fig{fig:dSumEt} (left), where we see that the {\tt PUPPI} algorithm noticeably improves the $\sum E_T$ resolution over {\tt PFlow} and {\tt PFlowCHS}.  While that fact that neither {\tt PFlow} nor {\tt PFlowCHS} are centered around zero is not an issue, the fact that {\tt PUPPI} is centered around zero supports the claim that applying {\tt PUPPI} produces a consistent event interpretation without the need for further pileup correction.

To compare the $\MET$ resolution, we look at the resolution of the $x$-component of $\MET$, shown in \Fig{fig:dSumEt} (right).  The relevance of this quantity to phenomenology is more directly seen, as this is one component of the $\MET$ vector.  Both the length and direction of the $\MET$ vector are important discriminating variables in many new physics searches so it is plausible that a small signal could be washed out by poor $\MET$ resolution or non-unity $\MET$ response.  We find that in our simplified set-up {\tt PUPPI} displays improvements over {\tt PFlow} and {\tt PFlowCHS}.  In fact, the resolution for {\tt PFlowCHS} degrades the resolution with respect to {\tt PFlow}.  This effect is due to the observation above that the partial removal of pileup interactions can lead to larger $\MET$ resolution.

For the pileup-reduced $\MET$ computations in CMS~\cite{CMS-PAS-JME-12-002}, it was found that the key component to reducing the pileup dependence on the $\MET$ resolution resulted from the identification and (indirect) removal of pileup jets from the $\MET$ calculation.  With {\tt PUPPI}, pileup jet removal is naturally built into the algorithm, thereby allowing for a simplified approach to pileup mitigation in $\MET$ related quantities. We expect that given the algorithm's flexibility in using experimental information, the improvement will persist in the full detector environment.

\section{Summary and Outlook}
\label{sec:summary}

In this paper, we have introduced a new algorithm, {\tt PUPPI}, for removing pileup contamination.
This method employs a {\it per particle} approach and improves the reconstruction of not only jet quantities, but also of event-wide observables like missing transverse energy. {\tt PUPPI} operates by using charged pileup to characterize the pileup in an event and then uses that knowledge to assign a weight to particles of unknown origin, like neutral hadrons or any particle in the forward region.  The weight is used to rescale the particle's four-momentum. The parameters of the algorithm are the size of the cone used to define neighboring particles $R_0$, the minimum distance cutoff $R_\text{min}$, the cut on the weights $w_\text{cut}$, and the cut on the rescaled transverse momentum $p_{T,\text{cut}}$.

By applying corrections at the particle level, before jet clustering, we can simultaneously perform pileup jet mitigation, and jet four-vector and shape corrections.  We have shown the improvement of {\tt PUPPI} over existing methods by studying jet $p_T$, mass, and missing transverse energy  over a wide range of jet $p_T$ and number of pileup interactions. Also, our method can be applied both in the central region of the detector (where tracking information is available) and in the forward region. 

In this work we have introduced the simplest form of the algorithm. However, many modifications and extensions are worth further exploration.  In particular, we have shown results for a single choice of metric, a particular weighting function, and a choice of how to use the weights.  Further modifications considered for the metric can include a combination of discrimination power from a selection of metrics into a common multivariate discriminant.  Preliminary studies with a boosted decision tree show modest improvements, although we leave it to future work to fully explore this avenue.

With regards to the particle weights, we have elected to allow fractional weights and chosen to use them to rescale four-vectors.  It is not obvious that a four-vector rescaling is the optimal procedure to implement.  As a simplification one could restrict weights to zero or one, in which case no rescaling is performed and particles are either kept or discarded.  Taking a step in the opposite direction, one could interpret the weights as probabilities that a particle should be kept in the event.  Given a probabilistic interpretation of weights, a natural approach would follow along the lines of Qjets~\cite{Ellis:2012sn,Kahawala:2013sba}, where a given event would yield many event interpretations with particles either kept or discarded according to their probabilistic weight.  All observables for a single event would then become distributions.  We leave this study for future work.

Though we frame our studies within a ``particle flow"-like set of inputs, it is not restricted only to inputs of this type.  If we consider as inputs calorimeter clusters rather than particles, we can still similarly compute the distance of tracks to a given calorimeter cluster, $i$, within the tracking volume.  Then for the forward region, we can consider nearby calorimeter clusters.  The challenge is to identify pure pileup clusters; we expect this can be achieved using tracks from the non-leading vertices.

This method may also be applied to heavy ion events, where energy densities of the underlying event are similar to the levels of pileup studied. In this case, however, all particles originate from the leading vertex.  One can use a modified version of {\tt PUPPI} in which the leading vertex constraint is not applied in the algorithm.

Finally, given the performance of {\tt PUPPI} on jet mass and $\MET$ we are optimistic that the {\tt PUPPI} algorithm will be useful in improving pileup mitigation of other jet and event shapes, and more generally in the identification of other physics objects.  For instance, given a pileup-corrected event, it is reasonable to expect that identifying isolated photons or leptons will be improved using a {\it per particle} weighting scheme.

\para While {\tt PUPPI} was developed, another particle level pileup removal method called SoftKiller~\cite{Cacciari:2014gra} has been proposed.  Preliminary comparisons indicate comparable performances~\cite{2014PUworkshop}.

\section*{Acknowledgments}

The authors are grateful to Jeff Berryhill, Matteo Cacciari, Dinko Ferencek, David Krohn, Andrew Larkoski, Filip Moortgat, Salvatore Rappoccio, Gavin Salam, Matthew Schwartz, Gregory Soyez, Jesse Thaler, and Lian-Tao Wang for useful discussions.  We would also like to thank the organizers Filip Moortgat, Gavin Salam, and Ariel Schwartzman, as well as the participants of Mitigation of Pileup Effects at the LHC workshop for many fruitful discussions.

DB is partly supported by the U.S. Department of Energy under cooperative research agreement Contract Number DE-FG02-05ER41360, by the LHC-Theory Initiative Graduate Fellowship, and by Istituto Nazionale di Fisica Nucleare (INFN) through the Bruno Rossi Fellowship, PH is supported by CERN, ML is supported by NSERC of Canada, and NT is supported by the Fermi Research Alliance, LLC under Contract No. De-AC02-07CH11359 with the United States Department of Energy and the LPC-CMS Distinguished Researcher program operated through FNAL.

\bibliographystyle{JHEP}
\bibliography{puppi}
\end{document}